# Title
Discovery of Log-Periodic Oscillations in Ultra-Quantum Topological Materials

# Authors
Huichao Wang,[1,2,3,†] Haiwen Liu,[4,†] Yanan Li,[1,2] Yongjie Liu,[5] Junfeng Wang,[5] Jun Liu,[6] Jiyan Dai,[3] Yong Wang,[6] Liang Li,[5] Jiaqiang Yan,[7] David Mandrus,[7,8] X. C. Xie,[1,2,9,*] and Jian Wang[1,2,9,*]

# Affiliations
[1]International Center for Quantum Materials, School of Physics, Peking University, Beijing 100871, China.
[2]Collaborative Innovation Center of Quantum Matter, Beijing 100871, China.
[3]Department of Applied Physics, The Hong Kong Polytechnic University, Kowloon, Hong Kong, China.
[4]Center for Advanced Quantum Studies, Department of Physics, Beijing Normal University, Beijing, 100875, China.
[5]Wuhan National High Magnetic Field Center, Huazhong University of Science and Technology, Wuhan 430074, China.
[6]Center of Electron Microscopy, State Key Laboratory of Silicon Materials, School of Materials Science and Engineering, Zhejiang University, Hangzhou, 310027, China.
[7]Materials Science and Technology Division, Oak Ridge National Laboratory, Oak Ridge, Tennessee 37831, USA.
[8]Department of Materials Science and Engineering, University of Tennessee, Knoxville, Tennessee 37996, USA.
[9]CAS Center for Excellence in Topological Quantum Computation, University of Chinese Academy of Sciences, Beijing 100190, China.
[*]Correspondence authors. Email: jianwangphysics@pku.edu.cn (J.W.); xcxie@pku.edu.cn (X.C.X).
[†]H.W. and H. L. contributed equally to this work.

# Abstract
Quantum oscillations are usually the manifestation of the underlying physical nature in condensed matter systems. Here we report a new type of log-periodic quantum oscillations in ultra-quantum three-dimensional topological materials. Beyond the quantum limit (QL), the log-periodic oscillations involving up to five oscillating cycles (5 peaks and 5 dips) are observed on the magnetoresistance (MR) of high quality single-crystal $ZrTe_5$, virtually showing the clearest feature of discrete scale invariance (DSI). Further theoretical analyses show that the two-body quasi-bound states can be responsible for the DSI feature. Our work provides a new perspective on the ground state of topological materials beyond the QL.

# Main Text
## Introduction
The DSI feature originates from the log-periodic corrections to scaling, and consequently observables are scale invariant only for certain geometrical scaling factors, reminiscent of fractal systems (*1*). In a scale-invariant system, the quantization effect can spontaneously break the continuous scale invariance down to DSI, which is a theoretically identified issue in quantum physics (*2*) but with rare manifestation in experiments. Governed by the Efimov equation, the DSI emerges as a distinctive feature of Efimov trimer bound states (*3-5*), which have been observed in cold atom experiments (*6-10*). The exotic scaling law and mathematical description of the Efimov effect can be further shared by the Efimov-like or Efimovian phenomena (*11*). In condensed matter physics, the DSI has been theoretically proposed to appear in quasi-bound states of massless Dirac fermions in the atomic collapse under supercritical Coulomb attraction (*12-14*). However, experimental evidence for the clear demonstration of the DSI has not been presented. The recently studied Dirac system with two types of carriers can satisfy the supercritical collapse condition for the appearance of DSI, providing a promising platform to search for this rare and important phenomenon in quantum physics. Besides, the potential link between the supercritical atomic collapse in topological systems and the appearance of DSI feature is also of particular interest.

In condensed matter physics, quantum oscillation revealed by magnetotransport investigation has been a powerful experimental technique to detect the underlying physics of solid-state systems. In the presence of a magnetic field (*B*), the Shubnikov-de Haas (SdH) oscillations showing a periodicity in $1/B$ can be usually observed at low temperatures and high magnetic fields for a clean single crystalline material (*15*). As a paradigm of Landau quantization of the energy levels, the SdH effect provides insights towards mapping the Fermi surface (*16*). Besides, in ring or cylinder structures, the Aharonov–Bohm (AB) and Altshuler-Aronov-Spivak (AAS) effects can also induce quantum oscillations in MR, where the oscillations are periodic in *B*. The observation of these effects offers an illustration of the quasi-particle quantum interference in mesoscopic systems (*17*). Thus, it would be interesting to explore the DSI behavior by magnetotransport measurements.

In this work, we report a new type of log-periodic quantum oscillations which are clearly demonstrated by the systematical magnetotransport results from different samples and different facilities with the maximum magnetic field up to 58 T. For the underlying physics, we find it cannot be understood in the scenarios of conventional quantum oscillations such as the SdH effect (even considering the Zeeman splitting) or other previously known mechanisms beyond the QL. On the other hand, the log-periodicity of the structures in the MR is reminiscent of the DSI behavior, which indicates that the system has a geometric series of length scales. Further theoretical derivations show that the Dirac fermions with supercritical Coulomb attraction can give rise to the two-body quasi-bound states with DSI feature. Our work provides a new perspective on the ground state of topological materials beyond the QL. More importantly, the discovery of a new type of quantum oscillations with the log*B* periodicity represents a new phenomenon beyond the Landau level physics, and this is the first time that the DSI is revealed by magnetotransport measurements in condensed matter systems. Besides, our work also indicates that the intriguing log-periodic oscillations are potentially universal in the topological materials with Coulomb attraction, which opens up a new direction to explore the DSI behaviour and the atomic collapse phenomenon.

**Results**
**Crystal characterizations**
$ZrTe_5$ crystallizes in a layered orthorhombic structure with the space group Cmcm (*18*). Along the *a* axis, Zr atoms and Te atoms are bonded as trigonal prismatic chains of "$ZrTe_3$", which are linked along the *c* axis via parallel zigzag chains of "$Te_2$". This forms one layer of $ZrTe_5$ in the ac plane and individual layers are coupled via van der Waals interactions along the *b* axis. The $ZrTe_5$ material was ever intensively investigated for the resistivity peak at certain temperatures and its SdH oscillations were reported to reveal the very tiny and anisotropic Fermi surface (*18-20*). Lately, the theoretical prediction of topologically nontrivial nature of the material has triggered a new wave of research boom (*21*). Based on the recent reports on the system from both theorists and experimentalists (*22-35*), it is known the system is extremely sensitive to the cell volume and thus the measured physical properties of $ZrTe_5$ are divergent in different samples modified by the growth condition (table S1).

ZrTe$_5$ single crystals used in this work were grown out of Te-flux method (*36*). The samples were well characterized by measuring x-ray single crystal diffraction, elemental analysis, electrical resistivity, and thermopower, confirming that our crystals are stoichiometric (*22*). Our energy dispersive spectroscopy results reveal an atomic ratio of the samples with Zr:Te ≈ 1:5 (table S2). We also used a FEI TITAN Cs-corrected cross-sectional STEM operating at 200 kV to further examine the crystalline nature of the ZrTe$_5$ sample. Figure 1A shows the atomic layer-by-layer high angle annular dark field (HAADF) STEM image of a typical ZrTe$_5$ sample, which demonstrates the high-quality nature. The deduced lattice constants of $a = 0.398$ nm and $b = 1.450$ nm (inset of Fig. 1A) are consistent with previous reports (*21-23*).

**Temperature dependence of resistivity and Hall traces**
Typical resistivity vs. temperature (*RT*) behavior of the ZrTe$_5$ crystals (s1 & s6) is shown in Fig. 1B. A crossover can be observed in the *RT* curves from metallic behavior above 200 K to a semiconducting-like upturn with saturation at low temperatures. The samples from the same batch show similar properties (fig. S1). The *RT* characteristic differs from those in most of the previous literature, in which the ZrTe$_5$ usually shows a sample dependent resistance peak at 60~150K. On the other hand, similar *RT* behavior of ZrTe$_5$ is also observed and reported by other research groups (*22,32*), and the absence of the resistance peak is attributed to the much smaller density of impurities and defects in the samples. This proposal appears to be consistent with our results because the Te deficiency can be largely reduced in our samples by using the self-Te flux method and further modifying the growth parameters (*22*).

Figure 1C shows the measured Hall traces obtained on s1, which indicates dominant hole carriers in the ZrTe$_5$ crystals. Based on the previous literature (*22-34*), the Hall behavior in ZrTe$_5$ shows dissimilarities in different samples and the carriers at low temperatures are reported to be holes, electrons or both types. For the samples with a resistance peak in the *RT* behavior, it is reported that the carriers change from dominated electrons at lower temperature to dominated holes at higher temperatures (*29*). The sign-change of Hall at the temperature where the peak appears is attributed to a proposal of Lifshitz transition. However, this would not happen in the samples without showing a resistance peak (*22,32*). The differences of the carriers in such two classes of ZrTe$_5$ samples are compared and clarified recently (*22*). Our Hall results are consistent with those of the samples without a resistance peak (*22,32*). Besides, most previous reports show the existence of hole carriers at low temperatures in ZrTe$_5$ crystals (*22,24,25,34*), which also support our observation in Hall measurements.

According to the analysis of measured magnetoresistivity and Hall resistivity, we obtain the conductivity tensor σ$_{xx}$ and σ$_{xy}$ (fig. S2). In a two-carrier model analysis of the σ$_{xx}$ and σ$_{xy}$ (*37*), the carrier density and mobility of our sample at selected temperatures are estimated and shown in Fig. 1D. The results indicate that the dominant hole carriers have a quite low density of $2.6\times10^{15}$ cm$^{-3}$ and a high mobility of about $3.9\times10^5$ cm$^2$/V/s at 2 K (*25,29*). In the ZrTe$_5$ system with a very low carrier density, the QL can be reached under a very small magnetic field (*31*). The estimated QL for our samples is about 0.2 T (see Methods), which offers an exciting playground to explore new physics in the ultra-quantum regime.

**Log-periodic MR oscillations**
Figure 2A shows the longitudinal MR behavior of s6 at low temperatures in a perpendicular magnetic field (*B//b* axis). We select the resistance at 5 T for a comparison of the MR results in different literature. In our samples, the R(5T)/R(0T) varies from 3 to 8, which is in the same order of magnitude with most of the previous publications (*22,23,30,33,34*). Nevertheless, it is found that the reported value of R(5T)/R(0T) at around 2 K can vary from 1.2 to 235 and the shape of the reported MR curves are also varied for different ZrTe$_5$ samples (all with *B//b* axis) (*22-34*).

In a semi-logarithmic scale, MR oscillations are observed that are superimposed on a large MR background. By computing the second derivative of the raw MR data at 2 K, the oscillations can be seen more clearly as shown in the inset of Fig. 2A. Two peaks at ~1.1 T and ~2.9 T are obvious. Extended measurements are further performed in an ultrahigh magnetic field up to 58 T. At 4.2 K, distinct MR oscillations are also observed in the small magnetic field regime and more oscillations appear above 3 T (Fig. 2B). To guide the eye, all the observable MR extrema are marked by the dash-dotted lines. The approximate equidistance of lines indicates that the oscillations are virtually periodic in the logarithmic magnetic field. The log-periodic law is more distinctly revealed after a background subtraction (see Methods and fig. S3). As shown in Fig. 2C, the novel oscillations are reproducible in different $ZrTe_5$ samples (s2, s7 and s9, s10, see fig. S4 and fig. S5). In the Fast Fourier Transform (FFT) analysis of the oscillation data ($\Delta R$ vs. log($B/B'$)), the sharp frequency ($F$) peak at $F \approx 2$ confirms the log$B$ period. Here $B'=1$ T is used to realize a dimensionless transformation and the full width at half maximum (FWHM) of the peak is respectively labeled in Fig. 2D.

As the temperature is increased, the log-periodic oscillations become weaker and gradually disappear one after the other (Fig. 3A). Nevertheless, it is surprising that the oscillations above 10 T could still survive at the temperature as high as 100 K (Fig. 3B). Figure 3C shows the temperature dependence of the log-periodic oscillations more clearly. Five oscillating cycles are observed at 4.2 K while only two cycles in large magnetic fields remain visible when the temperature is increased to 100 K, and the oscillation feature is gone at 150 K. The FFT results at the selected temperatures further reveal that the oscillations maintain the same log-period with the increasing temperature but disappear at 150 K (Fig. 3D).

We index the observed MR peak with an integer ($n$) and dip with a half ($n$-0.5), and the corresponding magnetic field is signed as $B_n$. The index plot for the oscillations at 4.2 K is shown in Fig. 4A. The behavior of $1/B_n$ vs. $n$ largely deviates the linear dependence (I-III in Fig. 4A), which is apparently different with the SdH effect even considering Zeeman splitting (see more details in Supplementary Materials). The linear dependence in the plot of $B_n$ vs $n$ is also absent (fig. S6). In a semi-logarithmic scale (IV-VI in Fig. 4), the linear dependence of log($B_n$) on $n$ further confirms the log-periodic property of the observed oscillations. In condensed matter systems, previously known quantum oscillations are mainly associated with either the Landau level physics (*16*) which is periodic in $1/B$ or the quantum interference effect (*17*) which is periodic in $B$. Our discovery of the log$B$ periodicity appreciably reveals a new class of quantum oscillations in condensed matter physics.

**Discussions and theoretical explanation**

The log-periodic structures in the MR data are reminiscent of the DSI behavior, which indicates that the system has a geometric series of length scales (*1*). Based on the Hall data (Fig. 1C), the high mobility hole is from the Dirac band with linear dispersion (*25,29*). Since the carriers are very dilute, the system can be simplified into a two-body problem. The charge impurity or electron from trivial band can act as a charge center to massless hole, and give rise to a long-range Coulomb attraction $V(\vec{R}) = \frac{-e^2}{4\pi\varepsilon_0 R}$ with $R$ denoting the distance (Fig. 4B). Because all terms are of the order $R^{-1}$, the massless Dirac Hamiltonian with Coulomb attraction remarkably obeys the scale invariance. After eliminating the angular wave function, the radial equation of each spinor element can be obtained, which is close to the Efimov equation (see more details in Supplementary Materials): $-\frac{d^2}{dR^2}u(R) + \frac{\kappa^2}{R^2}u(R) - \left(\frac{E}{\hbar v_F} + \frac{\alpha}{R}\right)^2 u(R) = 0$. Here $E$ is the energy, $\alpha = \frac{e^2}{4\pi\varepsilon_0 \hbar v_F}$ is the fine structure constant, $u(R)$ denotes the radial part of the spinor eigenfunction, and $\kappa = \pm 1, \pm 2 \ldots$ denotes the angular momentum index (*12*). The supercritical Coulomb attraction with $\alpha > |\kappa|$ (supercritical collapse condition) guarantees the formation of quasi-bound states (*12-14*), while for subcritical case $\alpha < |\kappa|$ the states are absent. In the following, we focus on the lowest angular momentum channel with $\kappa = \pm 1$. Thus, the radial momentum satisfies the formula $p_R^2 = \hbar^2 \left[ \left( \frac{E}{\hbar v_F} + \frac{\alpha}{R} \right)^2 - \frac{1}{R^2} \right]$. The semi-classical quantization $\int_{R_0}^{R_n} p \cdot dr = n\pi\hbar$ results in the DSI for the radius of the quasi-bound states $\frac{R_{n+1}}{R_n} = e^{\pi/s_0}$ with $s_0 = \sqrt{\alpha^2 - 1}$. The magnetic field introduces a new length scale $l_B = \sqrt{\hbar c/eB}$ and breaks the DSI of the quasi-bound states down to approximate DSI. The effect of magnetic field can be quantitatively analyzed by comparing the Landau level spacing $E_B = \sqrt{2}\hbar v_F/l_B$ with the Coulomb attraction $V(R_n) = \alpha \hbar v_F/R_n$ where $R_n$ indicates the most probable radius of the *n*-th quasi-bound states (see more details in Supplementary Materials). Under the magnetic fields with $E_B < V(R_n)$, the system still possesses approximate spherical symmetry. However, the magnetic field generates a repulsive interaction which breaks the large size states with $R_n > \sqrt{2s_0}l_B$ when $E_B$ exceeds the Coulomb attraction energy. When the magnetic field is enlarged, the energy of the *n*-th quasi-bound states approaches the Fermi energy at the magnetic field $B_n$. Figure 4C shows the numerical simulation of the spectrum under the magnetic field, in which $E_0$ and $L_0$ denote the cutoff energy scale and the cutoff length scale, respectively. $|E_n|$ denotes the binding energy of the two-body quasi-bound states without any magnetic field, and $B_n$ denotes the magnetic field value for the appearance of the *n*-th quasi-bound states around the Fermi energy (Fig. 4C). The resonant scattering process between the mobile carriers and the two-body quasi-bound states influences the transport property at the Fermi level, which leads to a log-periodic correction to the MR.

Based on T-matrix approximation, we obtain the log-periodic oscillating component of the MR: $\Delta \rho_{xx} = \frac{\rho_0 \sqrt{B[T]} \eta^2}{\sin^2\left(\frac{s_0}{2} \ln\left(\frac{B}{B_0}\right)\right) + \eta^2} - c\rho_0 \sqrt{B[T]}$ (see details in Supplementary Materials). Here, the first term denotes the log-periodic oscillations, and the second term indicates the background subtraction. The detailed derivation of the fitting formulas, and more discussions on the effect of Zeeman effect are given in Supplementary Materials. By using this formula, the observed log-periodic oscillations in different samples s6, s7 and s9 at 4.2K are quantitatively reproduced (black curves) as shown in Fig4D. Based on the fits, an averaging value of $s_0 \approx 5.4$ is obtained (see details in Supplementary Materials) and thus the Fermi velocity $v_F \approx 4.0 \times 10^5$ m/s can be deduced for the Dirac bands in ZrTe$_5$, which is very close to the results in previous literature (*25, 29, 31*).

Based on the FFT results of the log-periodic oscillations (Fig. 2D), the frequency peak at $F \approx 2.00$ indicates a period of $\log(B_n) \approx 0.50$ and naturally a main scale factor $\lambda = B_n/B_{n+1} \approx 3.16$ for the ZrTe$_5$ system. Considering the FWHM as an error bar, a reasonable $\lambda$ range [2.30, 4.00] is deduced for s6 at 4.2 K. The $\lambda$ range is about [2.38, 5.80] and [2.4, 5.30] for s7 and s9, respectively. For channel with $\kappa=1$, the theoretically estimated factor $\lambda$ locates in the range of (2.76, 4.06), and the range becomes border when considering the contribution from higher angular momentum channels, which is consistent with our experimental observations. Based on the model, the amplitude of the oscillations is proportional to the occupation number of the quasi-bound states, which satisfies $N=N_0(1-\exp(-\Delta E/k_B T))$. The temperature dependence of the oscillatory amplitude of the $n=1$ peak is shown in Fig. 4E. The data points (blue points) are from two samples of s6 (circles) and s10 (squares) to show more experimental details. The fitting (orange curve) parameter of the binding energy $\Delta E \sim 10.3$ meV indicates a disappearance temperature $T_d \sim 120$ K, which is also consistent with our experimental results. Thus, the log-periodic MR oscillations can be interpreted by the two-body quasi-bound states scenario. These results would inspire further theoretical investigations to give clear description of the DSI feature in this many-body system. For example, the screening effect from the continuous band sets a long-range cut-off for the quasi-bound states, and the lattice constant gives the short-range cut-off.

Direct estimation from the carrier density indicates the range with DSI approximately locates in the range of 0.4-60 nm, and correspondingly the magnetic field locates in the range of 0.2-150 T. The continuous band also contributes to the MR, rendering as the envelope of the MR data.

Moreover, the contribution of higher angular momentum channels to the MR can broaden the approximate DSI range. Typically, a log-log plot is the best way to show the DSI feature in a system. However, in the experimental observations, out of the log-periodic oscillations, other scattering mechanisms also contribute to the measured MR with a non-oscillating background. Thus, background subtraction procedure is necessary to extract the magneto-oscillations as generally analyzed in condensed matter physics (*16*). The procedure would have influence on the oscillating amplitude, but the process does not affect the periodicity of the oscillations. Thus, the signal of the DSI feature can be detected by the remarkable log-periodicity of the MR oscillations, which has been clearly demonstrated in our work by the FFT results (Fig. 2D and Fig. 3D) and the index plot with a semi-log scale (Fig. 4A). Lastly, there is certain deviation between the experimental data and fitting curves in the small field limit and the large field limit. The reason for this deviation is twofold. Firstly, the subtracted background at both ending fields is largely affected by the boundary condition. Thus, a small error exists in the characteristic $B_n$ for the oscillations at the boundary magnetic fields due to the influence of subtracted background, which could lead to the deviation of experimental data from the fitting curves. Secondly, in the large magnetic field limit, the theoretical fitting curves with the consideration of the Zeeman effect can be closer to experimental data. More discussions on the fitting are given in Supplementary Materials.

Other physical mechanisms may also exist in a 3D electronic system beyond the QL (*30,38-41*), such as the fractional Hall effect, a Wigner crystal and a density wave transition. However, the observed oscillations do not agree with the behaviors of these states (see more details in Supplementary Materials). For example, the deformation or reconstruction of the Fermi surface by density-wave transition commonly occurs at a certain value of magnetic field with the carrier density influenced largely. Then, a remarkable sharp transition can be expected in the MR at the critical magnetic field. However, in our observations, the magneto-resistance does not show any sharp transition. More importantly, these mechanisms do not possess DSI, while it is a remarkable feature of our experimental results revealed by the peculiar log-periodicity of the five oscillating cycles. The DSI can also exist in a system with fractal property in real space (*1*). However, our samples are high quality single crystals with no signature of real space fractal or strong disorder induced multi-fractal properties. Therefore, the observed DSI is very likely a striking aspect of the quasi-bound states with geometrical scaling. The three-body model for Efimov states can be excluded, due to the harsh requirement of resonant scattering condition (see more details in Supplementary Materials). Because of the ultralow carrier density in our $ZrTe_5$ samples, the absence of screening effect can give rise to the Coulomb attraction, and the small Fermi velocity in $ZrTe_5$ further guarantees the supercritical collapse condition, which in combination result in the two-body quasi-bound states with DSI.

In summary, a new type of quantum oscillations different with previously known 1/B periodic SdH effect (*16*) and B periodic AB or AAS effect (*17*) in condensed matter systems has been revealed by our magnetotransport results, which may shed light on the ground state of topological materials beyond the quantum limit. The discovery of the exotic log-periodic oscillations involving five oscillating cycles is a clear manifestation of the rarely observed DSI, which is of high general interest to several fields of physics. Relevant indications for the DSI of the quasi-bound states in solid state materials may be further detected by other techniques such as the magnetic susceptibilities measurements, thermal transport measurements, and the scanning tunneling spectroscopy. Besides, it will be interesting to extend the present study to a broad range of topological materials (*42*) with Coulomb attraction, as the distinctive log-periodic MR oscillations and the underlying DSI-featured spectrum could be universal.

*Note added*: After completion of this work, we became aware of a preprint, which addresses the quasi-bound states and the signature of broken continuous scale symmetry in graphene by scanning tunneling microscope (*43*).

**Materials and Methods**

**Sample growth** $ZrTe_5$ single crystals used in this work were grown out of Te-flux method. In a typical growth, Zr slug and Te shots in an atomic ratio of 1:49 were loaded into a 2 ml Canfield crucible set (*36*) and then sealed in a silica ampoule under vacuum. The sealed ampoule was heated to 1000 ℃ and kept for 12 h to homogenize the melt, furnace cooled to 650 ℃, and then cooled down to 460 ℃ in 60 h. $ZrTe_5$ crystals were isolated from Te flux by centrifuging at 460 ℃. Typical $ZrTe_5$ crystals are about 10-20 mm long with the other two dimensions in the range of 0.01-0.4 mm.

**Crystal characterization** The samples were well characterized by measuring x-ray single crystal diffraction, elemental analysis, electrical resistivity, and thermopower, confirming that our crystals are stoichiometric (*22*). Our energy dispersive spectroscopy results reveal an atomic ratio of the samples with Zr:Te ≈ 1:5 (table S2). We also used a FEI TITAN Cs-corrected cross-sectional STEM operating at 200 kV to further examine the crystalline nature of the $ZrTe_5$ sample.

**Transport measurement** Electrical transport measurements in this work were conducted in three measurement systems, a Physics Property Measurement System (PPMS) from Quantum Design for the low temperature and static magnetic field measurements, a pulsed high magnetic field facility at Wuhan National High Magnetic Field Center (China) and a Dilution Refrigerator MNK126-450 system with static magnetic fields for an ultralow temperature environment. Results from different systems and different samples are reproducible and consistent. Standard four (six)-probe method was used for measuring resistivity (resistivity and Hall trace) with the excitation current always flowing along the *a* axis of $ZrTe_5$ in our electrical transport measurements.

**Calculation of Quantum limit (QL)** When all the carriers are confined to the lowest Landau level, the QL is reached. In recent work on $ZrTe_5$ crystals grown via chemical vapor transport using iodine as the transport agent, the carrier density is reported to be $10^{17}$-$10^{18}$ /cm$^3$ with a Shubnikov-de Haas (SdH) period of 3-5 T, which means that the QL can be reached in 3-5 T (*23,24*). By using Te-flux method, a lower magnetic field (about 1 T) is needed to drive the compound into its QL (*31*). In our growth conditions, the $ZrTe_5$ crystals have the desired stoichiometry and show very low carrier densities (*22*). According to the Onsager relation, our high quality $ZrTe_5$ samples with the much lower densities should show a smaller SdH period and simultaneously the critical field when the system enters the QL is smaller. We usually judge whether a system enters the QL by analyzing its SdH effect. In our samples, it is hard to extract the SdH oscillations which are merged into the sharp increase of MR around 0 T. However, we could estimate the QL magnetic field $B_c$ for our $ZrTe_5$ crystal based on the carrier density. The critical field at which the system enters the QL field can be estimated by the formula (*44*)

$B_c = \left(2\pi^4 n^2 \frac{\sqrt{m_a m_c}}{m_b}\right)^{1/3} \left(\frac{\hbar}{e}\right) \approx 3.8 \times 10^{-11} n^{2/3} \left(\frac{\sqrt{m_a m_c}}{m_b}\right)^{1/3}$, where the carrier density is in cm$^{-3}$, $m_a$, $m_c$ are masses perpendicular to the magnetic field, and $m_b$ is the mass along the field.

The very low carrier density and the strong anisotropic property (fig. S7) of our bulk $ZrTe_5$ crystals indicate a very small value of the critical field. Assuming the anisotropy of the carrier mass is constant for a specific material, one obtains the critical field $B_c \propto n^{2/3}$. In our calculation, the anisotropic masses for carriers reported in previous literatures are referred. It is estimated that the QL magnetic field for our crystals is about 0.2 T, which is rather small. Thus, the observed oscillations are beyond the QL, which also excludes the SdH effect as the underlying mechanism. Besides, the investigation on the ground state of a 3D electronic gas system beyond the QL is a long-standing research subject (*30,38-41*). Our work also provides an exciting playground to explore new physics beyond the QL.

**Data analysis** Oscillations in the raw MR data are clearly visible although the MR background is rather large. After subtracting a smooth background for clarity, the MR oscillations become more apparent and are in good agreement with the structures in the original MR data. To be more rigorous, we demonstrate here in details that how we extract the MR oscillations by different methods in the work to confirm that the oscillations are intrinsic.

In the inset of Fig. 2A in the main text, the oscillations are obtained by computing the second derivative for the measured raw data. This is a very convincing and undisputed method, which is usually used to pick up the maximum/minimum. By comparing the oscillations in the derivative result with that in the raw data, we could observe their correspondence and consistency. The oscillations in the static magnetic field measurements show linear dependence for $\log B_n$ vs $n$ (fig. S8) as in the Fig. 4A in the main text. The method frequently used to produce a background is doing polynomial fitting, which is used in our work for the creation of Supplemental Fig. S3C. The original raw data are shown in figs. S3A and 3B. After subtracting a sixth order polynomial from the raw data, we obtain the results shown in fig. S3C. In fact, the oscillations could be obtained whenever a $5^{th}/6^{th}/7^{th}/8^{th}$ polynomial is subtracted, and we could also observe the correspondence and consistency of the oscillations in fig. 3C with the oscillations on the raw data (figs. S3A and S3B).

For the MR in an ultrahigh magnetic field up to 58 T, the whole oscillations could not be shown clearly in the second derivative results due to the large amplitude changes. Meanwhile, a polynomial fitting could not produce a reasonable background curve in the large magnetic field regime. In this case, a smooth background is produced by smoothing the raw data, as shown in fig. S2. The black points are the raw data and the red line is the produced background. The background at the ends shows tiny anomaly due to unsuitable boundary conditions and we always discard the data at the boundaries for further analysis. After subtracting the background from the original data, distinct log-periodic MR oscillations are obtained. The obtained oscillations are also consistent with the oscillations in the raw data and the second derivative results.

In summary, the exotic MR oscillations periodic in logarithmic B are reproducible in different samples, even though the data are measured in different systems and analyzed by different methods. The results demonstrate that the oscillations are intrinsic properties of the high quality $ZrTe_5$ crystals.

**Supplementary Materials**
Supplementary material for this article is available online.
table S1. A brief review of the results of $ZrTe_5$.
table S2. A summary of the energy dispersive spectroscopy results on our $ZrTe_5$ crystals.
fig. S1. Resistivity vs. temperature behavior of $ZrTe_5$ crystals.
fig. S2. Hall analysis in a two-carrier model.
fig. S3. Background produced by smoothing the raw magnetoresistance data.
fig. S4. Log-Periodic MR oscillations in sample 2 (s2).
fig. S5. Log-periodic oscillations in ZrTe5 (s10).
fig. S6. The MR oscillations periodic in $\log B$ and not in $B$ or $1/B$.
fig. S7. Strong anisotropy of the bulk $ZrTe_5$.
fig. S8. $\log B_n$ vs. $n$ for the oscillations in s6 measured at the static magnetic field.
Supplementary Discussions on Other Physical Mechanisms
Supplemental Notes on the Theoretical Details
References (*44-56*)

**References and Notes**
1. D. Sornette, Discrete-scale invariance and complex dimensions, *Phys. Rep.* **297**, 239-270 (1998).
2. L. D. Landau, E. M. Lifshitz, *Quantum Mechanics: Non-relativistic Theory*, 3rd ed. (Perganon Press, Oxford, U.K., 1977).

**Acknowledgments**
We acknowledge Yanzhao Liu, Jinguang Cheng, Xi Lin, Ruoxi Zhang, Ziquan Lin for the help in related experiments and Haizhou Lu, Fa Wang, Qizhong Zhu for valuable discussions. **Funding:** This work was financially supported by the National Basic Research Program of China (Grant No. 2018YFA0305604, No. 2013CB934600, No. 2017YFA0303302 and No. 2015CB921102), the Research Fund for the Doctoral Program of Higher Education of China(20130001110003), the National Natural Science Foundation of China under Grants No. 11774008, No. 11534001, No. 11674028, No. 11504008, No. 51390474 and No. 11234011, the Open Project Program of the Pulsed High Magnetic Field Facility (Grant No. PHMFF2015002) at the Huazhong University of Science and Technology, and the Key Research Program of the Chinese Academy of Sciences (Grant No. XDPB08-1). J.-Q. Y. is supported by the US Department of Energy, Office of Science, Basic Energy Sciences, Materials Sciences and Engineering Division. D.G.M. was supported by the Gordon and Betty Moore Foundation's EPiQS Initiative through Grant GBMF4416. Huichao Wang acknowledges the Postdoctoral Fellowships Scheme of the Hong Kong Polytechnic University (Project No. 1-YW0T). **Author contributions:** J.W. conceived and instructed the experiments. H.W. performed the experiments and analyzed the experimental results. H.L. and X.C.X. proposed and developed the theoretical model. J.Y and D.M. synthesized the samples. Y. Li, Y. Liu, Junfeng Wang and L.L. helped in the transport measurements. J.L. and Y.W. performed the TEM characterization. H.W, H.L. and J. W. wrote the paper with the assistance of X.C.X., J.Y. and D.M. **Competing interests:** The authors declare that they have no competing interests. **Data and materials availability:** All data needed to evaluate the conclusions of the paper are present in the paper and/or the Supplementary Materials. Additional data related to this paper may be requested from the authors.

**Figures and Tables**

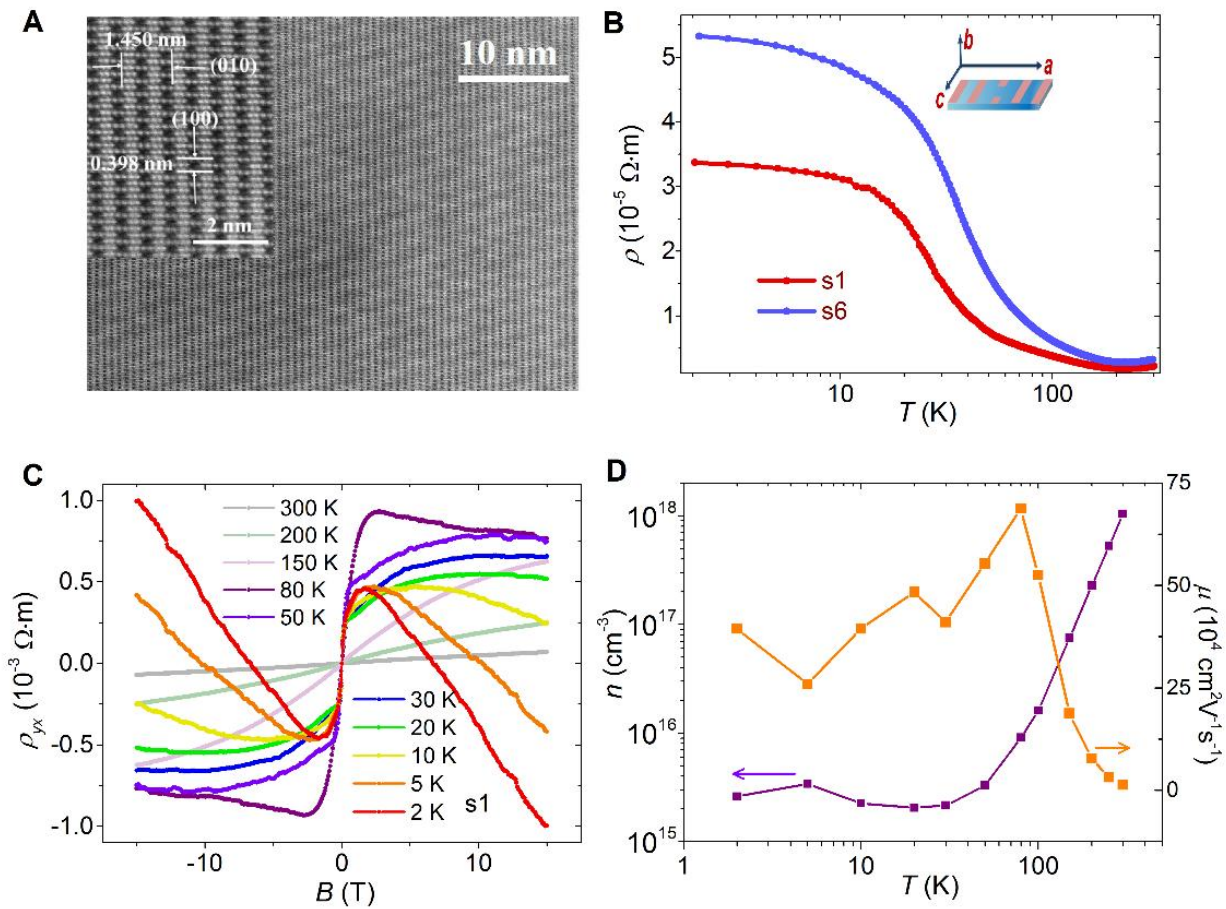

**Fig.1. Characterization of high quality ZrTe$_5$ crystal.** (**A**) HAADF STEM image of a typical ZrTe$_5$ sample. Scale bar represents 10 nm. Inset is the enlarged image showing atomic resolution with a scale bar of 2 nm. (**B**) *RT* characteristic of ZrTe$_5$. Inset shows the schematic for electrical transport measurements. (**C**) Hall traces of s1 vs. *B* at selected temperatures from 2 K to 300 K. (**D**) Temperature dependence of the estimated mobility and carrier density of the dominant holes in ZrTe$_5$ by analyzing the Hall data with a two-carrier model.

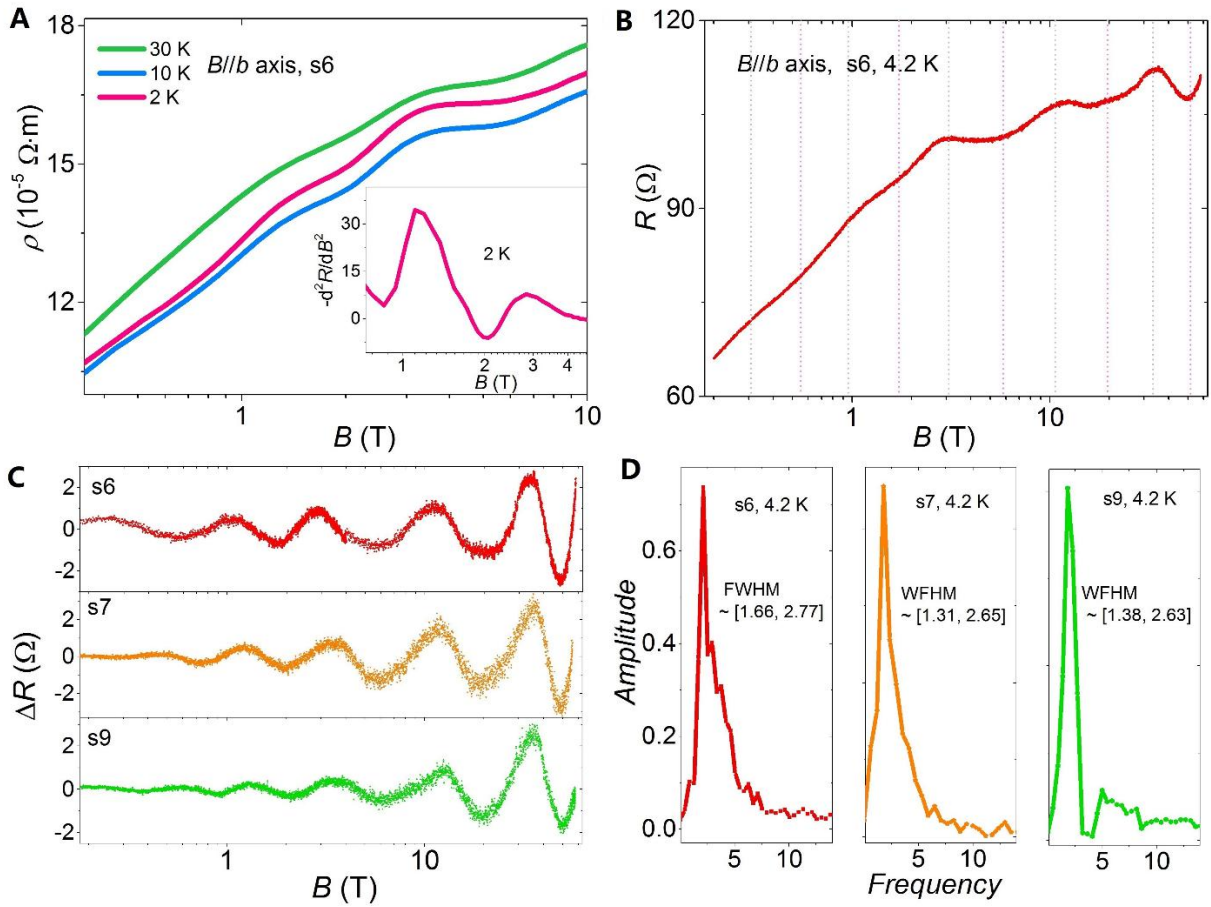

**Fig. 2. Log-periodic MR oscillations in ZrTe$_5$.** (**A**) Resistivity of s6 vs. $B$ at low temperatures in a static perpendicular magnetic field. Inset is the second derivative result of MR data at 2 K. (**B**) MR behavior of s6 in an ultrahigh magnetic field up to 58 T at 4.2 K. Dashed lines serve as guides to the eye. (**C**) Extracted MR oscillations in ZrTe$_5$ samples (s6, s7 and s9) at 4.2 K. (**D**) FFT results for the log-periodic oscillation data in the form of $\Delta R$ vs. $\log(B/B')$ with $B'=1$ T.

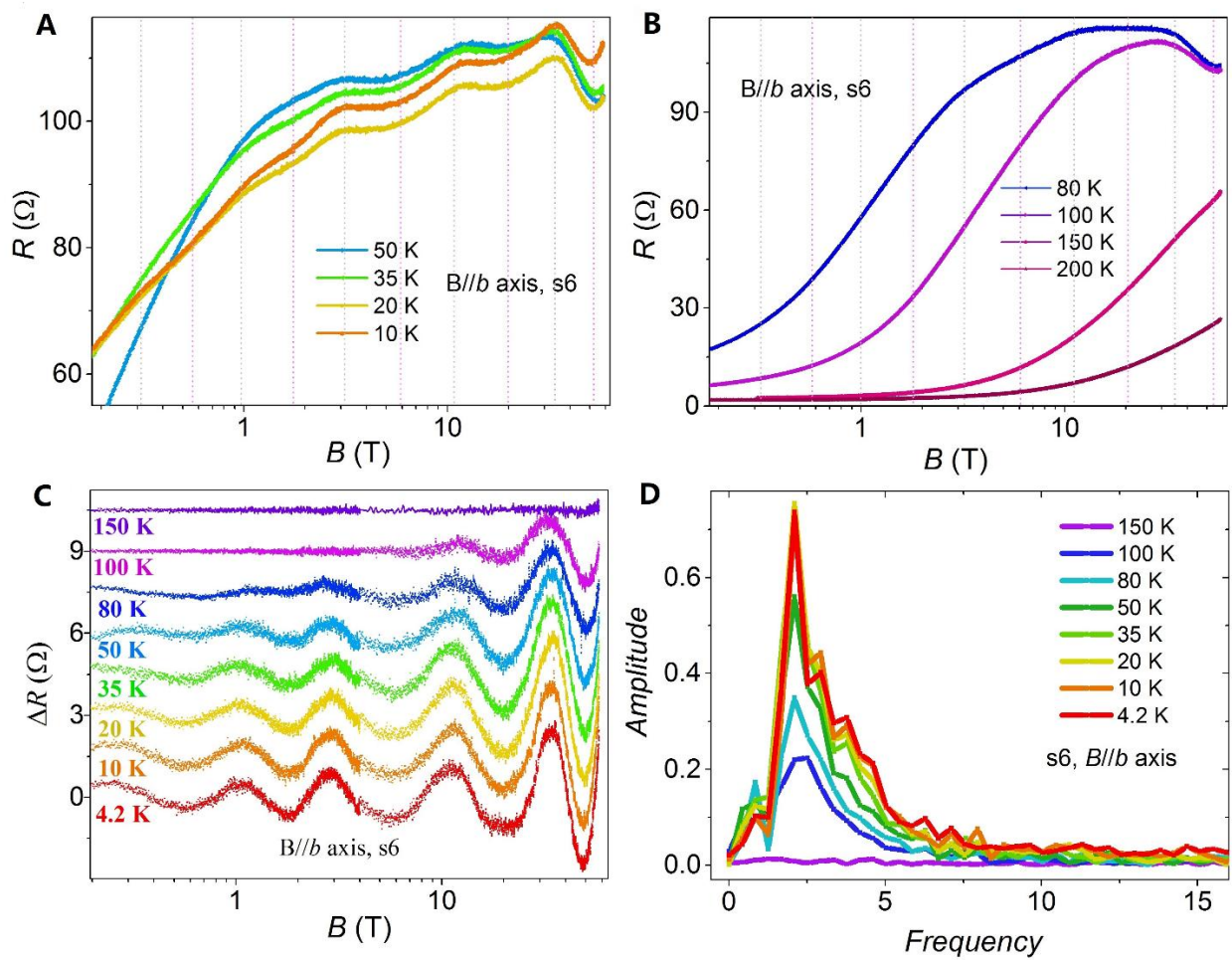

**Fig. 3. Log-periodic MR oscillations at different temperatures.** (**A**) MR behavior of s6 at relatively low temperatures. (**B**) MR behavior of s6 at relatively high temperatures. (**C**) MR oscillations in s6 after subtracting a smooth background from the raw data. Dashed lines serve as guides to the eye. (**D**) FFT results for the MR oscillations at different temperatures in the form of $\Delta R$ vs. $\log(B/B')$ with $B'=1$ T.

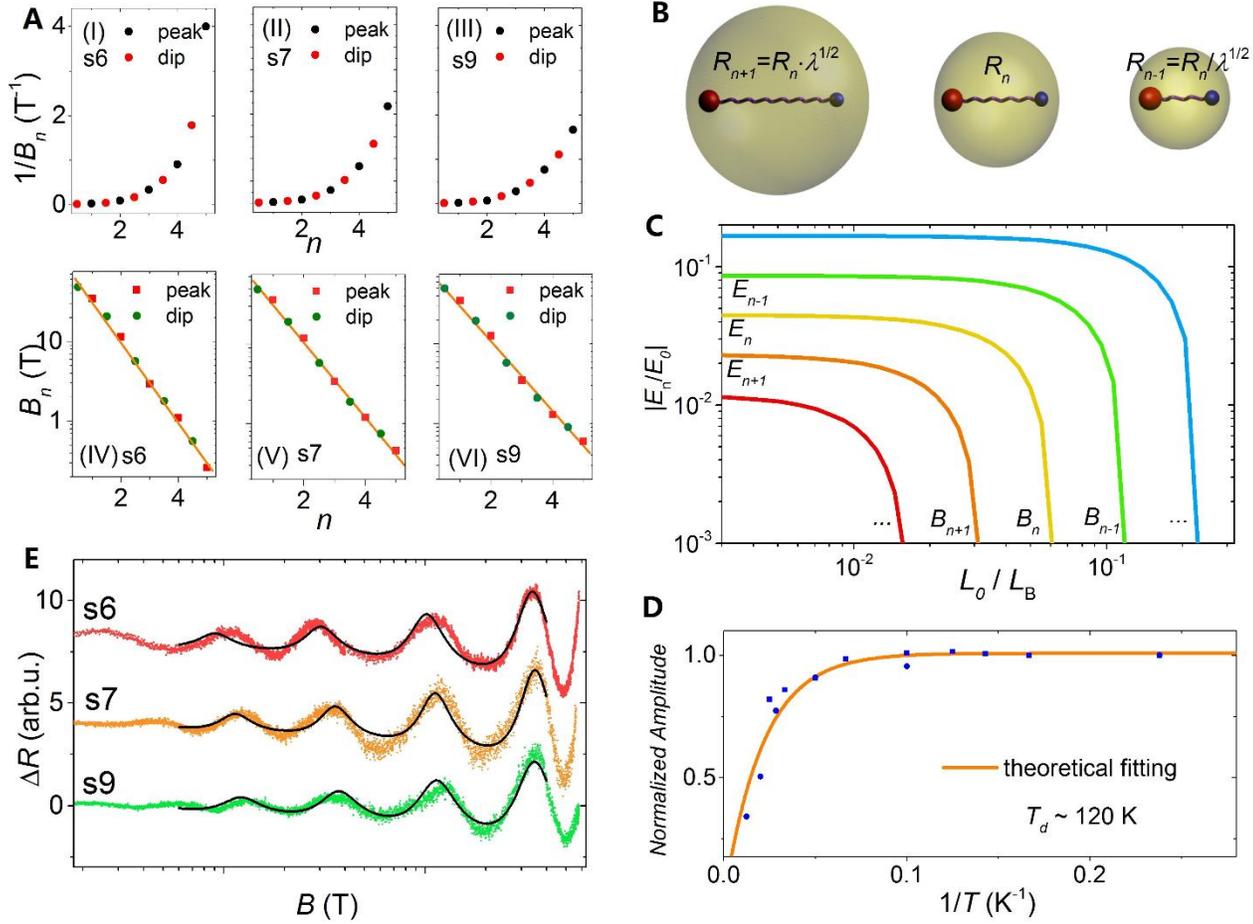

**Fig. 4. DSI in ultra-quantum ZrTe$_5$.** (**A**) Index plot for the log-periodic oscillations. (**B**) Schematic of the two-body quasi-bound states composed of a Dirac-type massless hole and charged center via Coulomb attraction. (**C**) Normalized binding energy of the quasi-bound states under the magnetic field. $E_0$ and $L_0$ denote the cutoff scale. (**D**) Quantitative fitting (black curves) of the log-periodic oscillations in ZrTe$_5$. (**E**) Temperature dependence of the normalized amplitude of the oscillation peak $n=1$. The fitting (orange curve) of the oscillation data (blue points) indicates a disappearance temperature $T_d$ consistent with the experimental results.

Huichao Wang,[1,2,3,†] Haiwen Liu,[4,†] Yanan Li,[1,2] Yongjie Liu,[5] Junfeng Wang,[5] Jun Liu,[6] Jiyan Dai,[3] Yong Wang,[6] Liang Li,[5] Jiaqiang Yan,[7] David Mandrus,[7,8] X. C. Xie,[1,2,9,*] and Jian Wang[1,2,9,*]

[1]International Center for Quantum Materials, School of Physics, Peking University, Beijing 100871, China.
[2]Collaborative Innovation Center of Quantum Matter, Beijing 100871, China.
[3]Department of Applied Physics, The Hong Kong Polytechnic University, Kowloon, Hong Kong, China.
[4]Center for Advanced Quantum Studies, Department of Physics, Beijing Normal University, Beijing, 100875, China.
[5]Wuhan National High Magnetic Field Center, Huazhong University of Science and Technology, Wuhan 430074, China.
[6]Center of Electron Microscopy, State Key Laboratory of Silicon Materials, School of Materials Science and Engineering, Zhejiang University, Hangzhou, 310027, China.
[7]Materials Science and Technology Division, Oak Ridge National Laboratory, Oak Ridge, Tennessee 37831, USA.
[8]Department of Materials Science and Engineering, University of Tennessee, Knoxville, Tennessee 37996, USA.
[9]CAS Center for Excellence in Topological Quantum Computation, University of Chinese Academy of Sciences, Beijing 100190, China.
[*]Correspondence authors. Email: jianwangphysics@pku.edu.cn (J.W.); xcxie@pku.edu.cn (X.C.X).

This file includes:
- table S1. A brief review of the results of ZrTe$_5$.
- table S2. A summary of the energy dispersive spectroscopy results on our ZrTe$_5$ crystals.
- fig. S1. Resistivity vs. temperature behavior of ZrTe$_5$ crystals.
- fig. S2. Hall analysis in a two-carrier model.
- fig. S3. Background produced by smoothing the raw magnetoresistance data.
- fig. S4. Log-Periodic MR oscillations in sample 2 (s2).
- fig. S5. Log-periodic oscillations in ZrTe5 (s10).
- fig. S6. The MR oscillations periodic in log$B$ and not in $B$ or $1/B$.
- fig. S7. Strong anisotropy of the bulk ZrTe$_5$.
- fig. S8. Log$B_n$ vs. $n$ for the oscillations in s6 measured at the static magnetic field.
- Supplementary Discussions on Other Physical Mechanisms
- Supplemental Notes on the Theoretical Details
- References (*44-56*)

# I. Supplemental Figures and Tables

**table S1. A brief review of the results of ZrTe$_5$**. Based on the recent intensive research from both theorists and experimentalists, it is known the ZrTe$_5$ material is extremely sensitive to the cell volume and thus the measured physical properties are divergent, which closely relates to the growth condition of the samples. The table is for a comparison of our results to those in previous literature and some omissions of the references are probably inevitable. The CVT method means the chemical vapor transport using iodine as the transport agent. We select the resistance at 5 T for the comparison of the MR effect. The summary of R(5T)/R(0T) and Hall information is for the condition when the magnetic field is along *b* axis. The blank is due to the lack of related information. The consistent properties of our samples with previous reports are marked as red color in bold font.

| Growth | Resistance Peak | R(5T)/R(0T) | Hall and Carriers Information | References |
|---|---|---|---|---|
| **Te-flux** | **< 2 K** | **3-6 (2 K)** <br> **~8 (0.3 K)** | **holes dominate (2 K-300 K)** | **The samples in this manuscript.** |
| Te-flux | 60 K | 235 (~5 K) | **holes from Dirac band dominant (20 K) (ARPES)** | Nat. Phys. **12**, 550-554 (2016) (*25*) |
| CVT | 145 K | 12 (2 K) | electrons dominate with holes (2 K-20 K) | Phys. Rev. B 93, 115414 (2016) (*24*) |
| CVT | 146 K | **10 (2 K)** | electrons dominate (2 K) | Phys. Rev. B 96, 121401(R) (2017) (*33*) |
| CVT | 138 K | **7 (2 K)** | electrons dominate (2 K-80 K), holes dominate (>120 K) | Nat. Commun. 7, 12516 (2016) (*23*) |
| CVT | 170 K | **7.5 (0.3 K)** | electrons dominate (0.3 K-10 K) | Sci. Rep. 6, 35357 (2016) (*30*) |
| Te-flux | 60 K | | electrons dominate (5 K-70 K), holes dominate (>70 K) | New J. Phys. 19015005(2017) (*29*) |
| CVT | **< 2 K** | **4 (2 K)** | **holes dominate (2 K-300 K)** | Sci. Rep. 7, 40327 (2017) |
| CVT | 132 K | 15 (0.3 K) | | Phys. Rev. Lett 118, 206601 (2017) |
| CVT | > 300 K | **1.2 (1.5 K)** | **holes dominate (1.5 K-300 K)** | Phys. Rev. B 95, 035420 (2017) (*34*) |
| CVT | 130 K | 32 (~2 K) | electrons dominate (<130 K) holes dominate (>130 K) | arXiv:1611.06370 (*22*) |
| Te-flux | < 2 K | **10 (~2 K)** | **holes dominate (2 K-300 K)** | arXiv:1611.06370 (*22*) |
| | **< 5 K** | | **holes dominate (17 K)** | Nat. Phys. 14, 451–455 (2018) (*32*) |
| Te-flux | 60 K | | **Very small quantum limit (1 T)** | Phys. Rev. Lett 115, 176404 (2015) (*31*); <br> Phys. Rev. B 92, 075107 (2015) |
| CVT | | | **Very small quantum limit (0.5 T)** | Phys. Rev. B 96, 041101 (2017) (*35*) |
| CVT | | | **holes (24 K) (ARPES)** | Phys. Rev. X 6, 021017 (2016) |
| CVT | 135 K | | electrons (< 135 K); holes (> 135 K) (ARPES) | Nat. Commun. 8, 15512 (2017) |
| CVT | | | holes (210 K) (ARPES) | Phys. Rev. Lett 117, 237601 (2016) |
| CVT | 135 K | | electrons dominate (< ~135 K); holes dominate (> ~135 K) | Journal of Crystal Growth 457, 250-254 (2017) (*27*) |
| CVT | 135 K | | electrons dominate (< ~135 K); holes dominate (> ~135 K) | J. Phys. Condens. Matter 16, L359-L365 (2004) (*26*) |
| CVT | 145 K | | | Phys. Rev. B 60, 7816-7819 (1999) (*19*) |
| CVT | 150 K | | | J. Phys. Soc. Jpn. 49, 839-840 (1980) (*18*) |

**table S2. A summary of the energy dispersive spectroscopy results on our ZrTe$_5$ crystals.**

| No. | Zr (Atomic ratio) | Te (Atomic ratio) |
|---|---|---|
| #1 | 16.50 | 83.50 |
| #2 | 16.71 | 83.29 |
| #3 | 16.87 | 83.13 |
| #4 | 17.10 | 82.90 |
| #5 | 17.16 | 82.84 |
| #6 | 16.16 | 83.84 |
| Average | 16.75 | 83.25 |
| Atomic ratio of Zr:Te = 16.75:83.25 ≈ 1:4.97 | | |

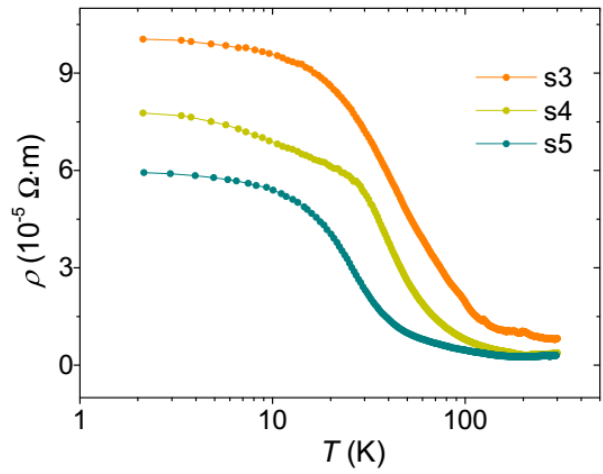

**fig. S1. Resistivity vs. temperature behavior of ZrTe$_5$ crystals.**

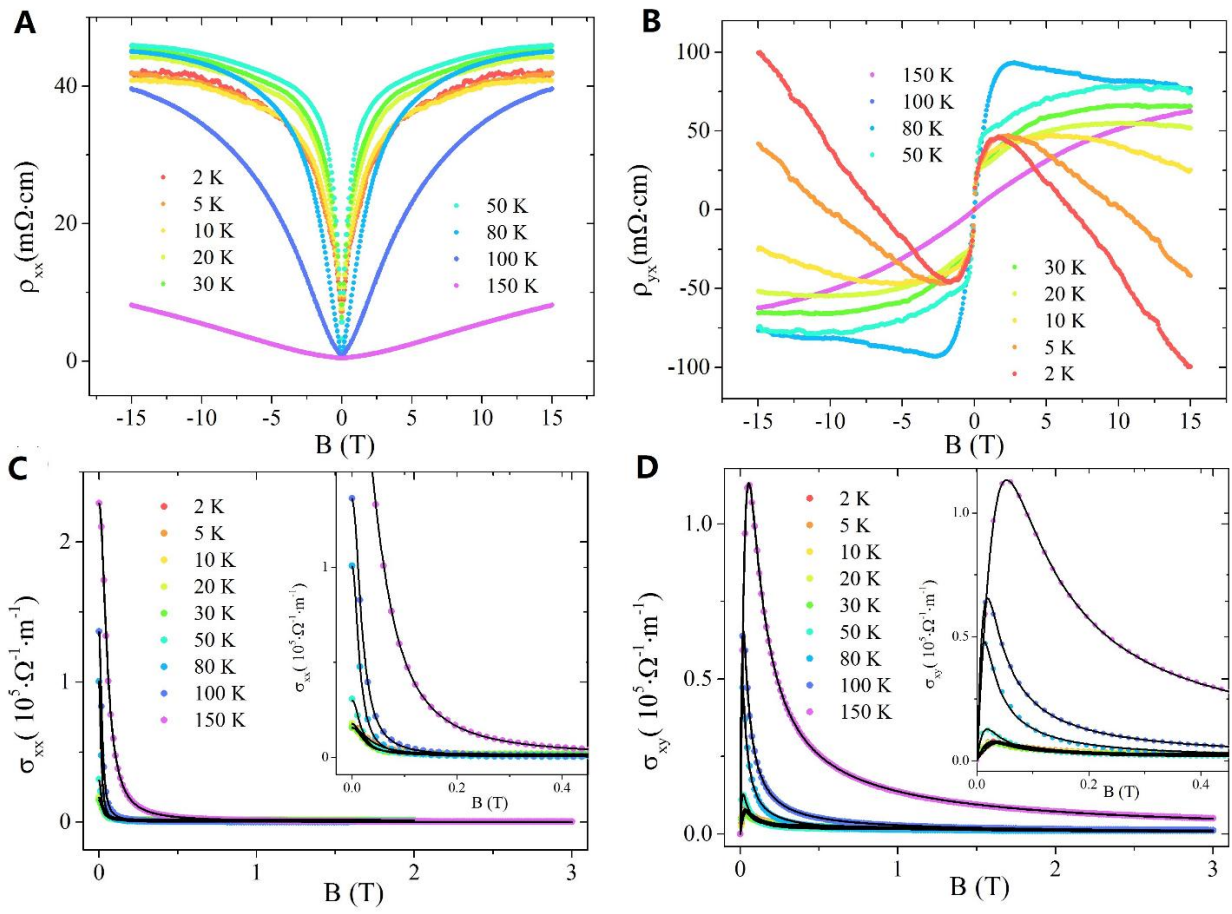

**fig. S2. Hall analysis in a two-carrier model.** (**A**) and (**B**) are the experimentally measured resistivity $\rho_{xx}$ and $\rho_{xy}$, respectively. (**C**) and (**D**) are the fits on the deduced conductivity $\sigma_{xx}$ and $\sigma_{xy}$, respectively. Black lines are the theoretical fits based on a two-carrier model.

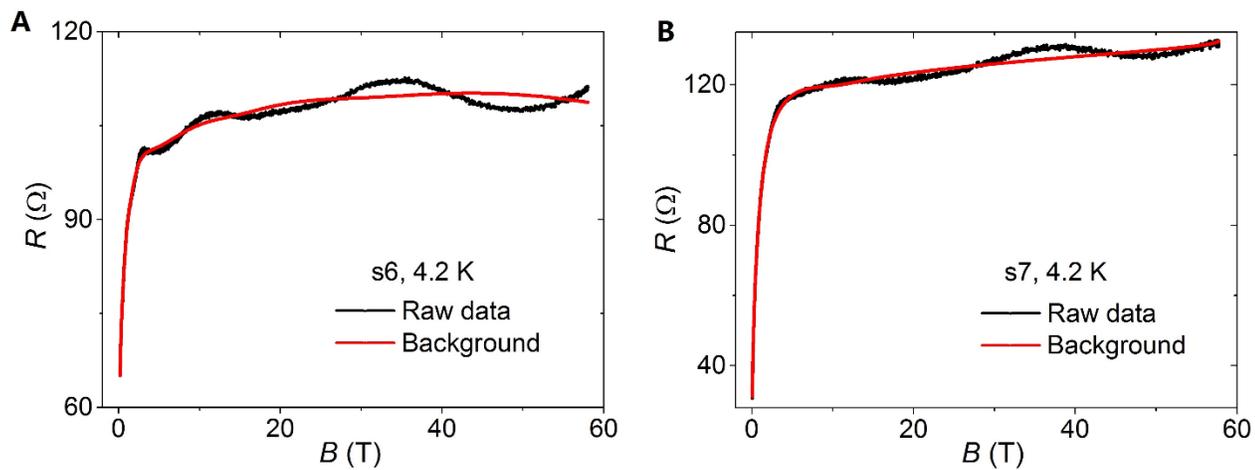

**fig. S3. Background produced by smoothing the raw magnetoresistance (MR) data.** The background and the raw MR data in (**A**) s6 and (**B**) s7 are presented by the red line and the black points, respectively.

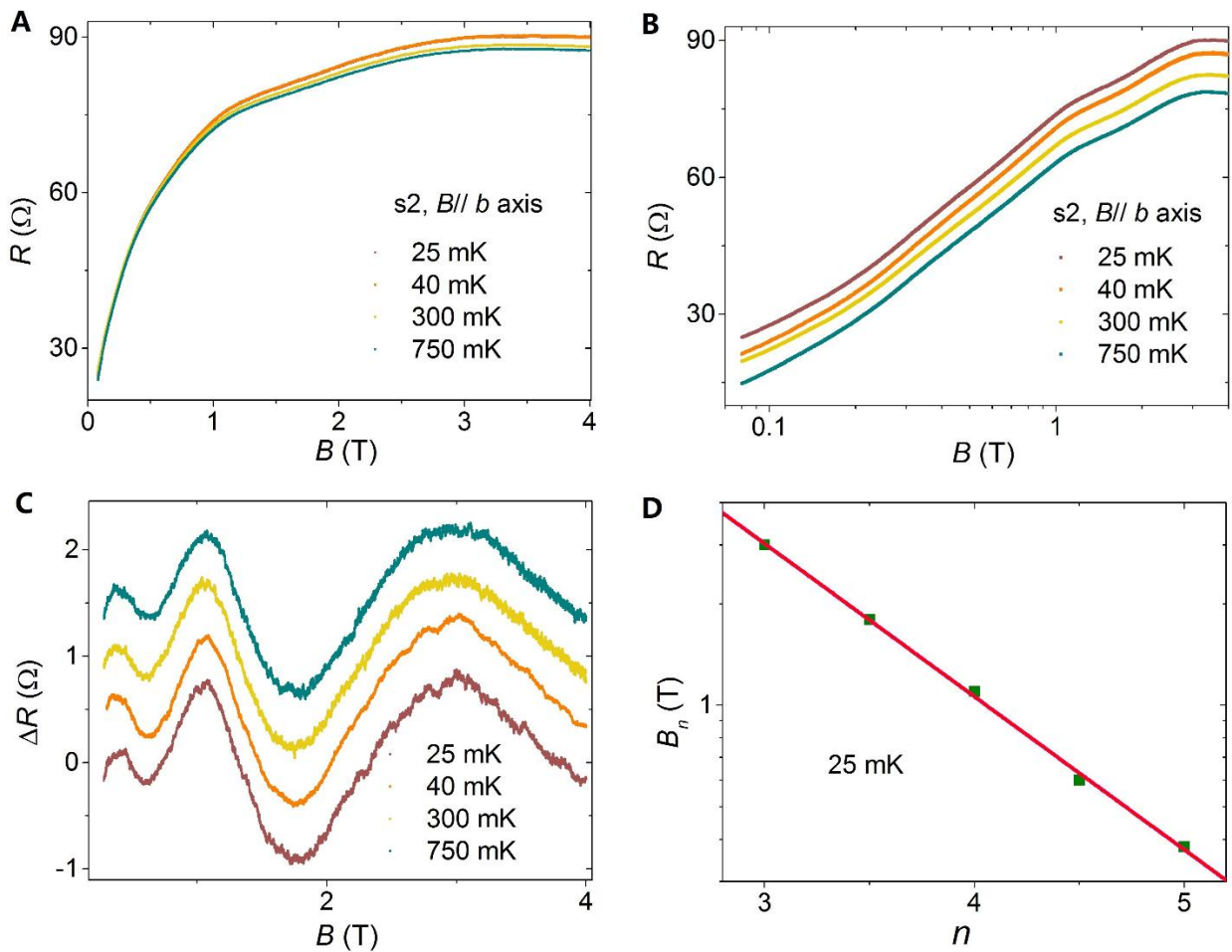

**fig. S4. Log-Periodic MR oscillations in sample 2 (s2).** (**A**) MR curves of s2 for 0-4 T in an ultralow temperature environment. (**B**) A semi-logarithmic scale is used and data above 25 mK are shifted for clarity of the MR oscillations in s2. (**C**) Extracted MR oscillations by subtracting a polynomial background from the raw MR data. (**D**) Linear dependence of $\log(B_n)$ on $n$ indicating a $\log B$ period of the MR oscillations.

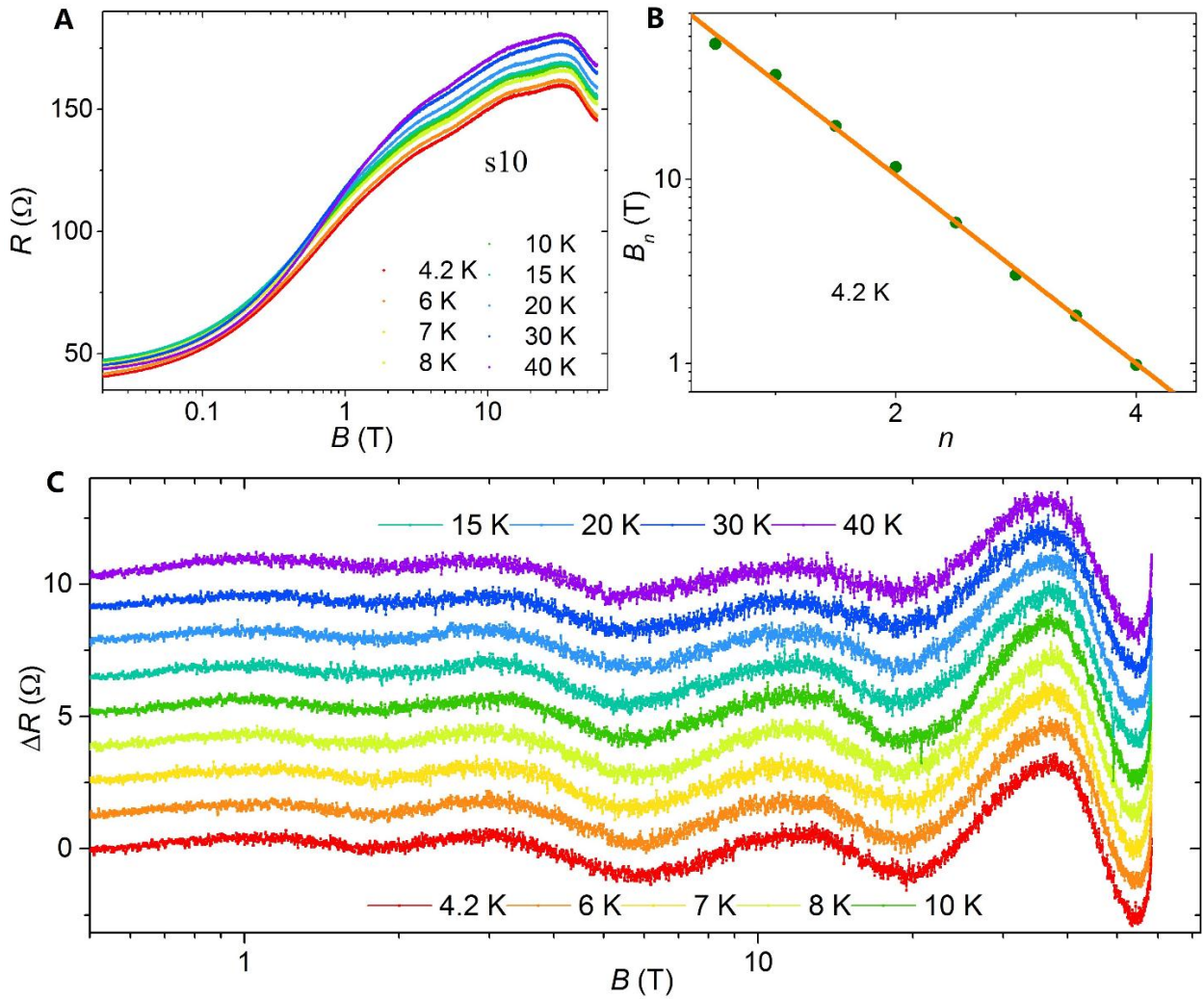

**fig. S5 Log-periodic oscillations in ZrTe$_5$ (s10).** (**A**) MR curves of s10 at selected temperatures in an ultrahigh magnetic field up to 58 T. (**B**) Linear dependence of $\log(B_n)$ on $n$ indicating a $\log B$ period of the MR oscillations. (**C**) Extracted MR oscillations from the raw MR data.

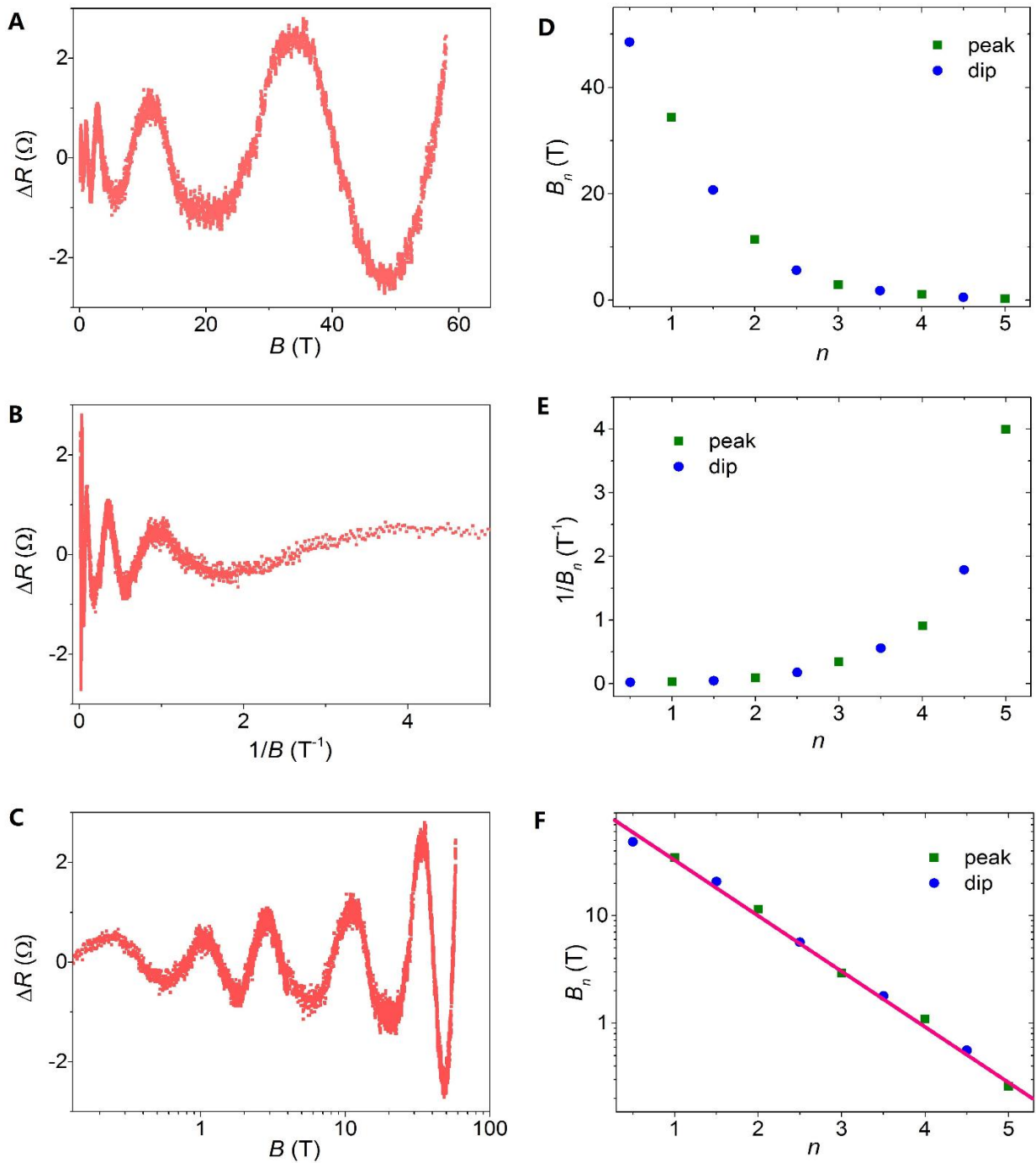

**fig. S6. The MR oscillations periodic in log$B$ and not in $B$ or $1/B$.** (A)-(C) Background subtracted data (in s6 at 4.2 K) plotted as a function of $B$, $1/B$ and log$B$. (D)-(F) Index dependence of $B_n$, $1/B_n$ and $\log(B_n)$. It is apparent that the oscillations are periodic only when the magnetic field is shown in a logarithmic scale.

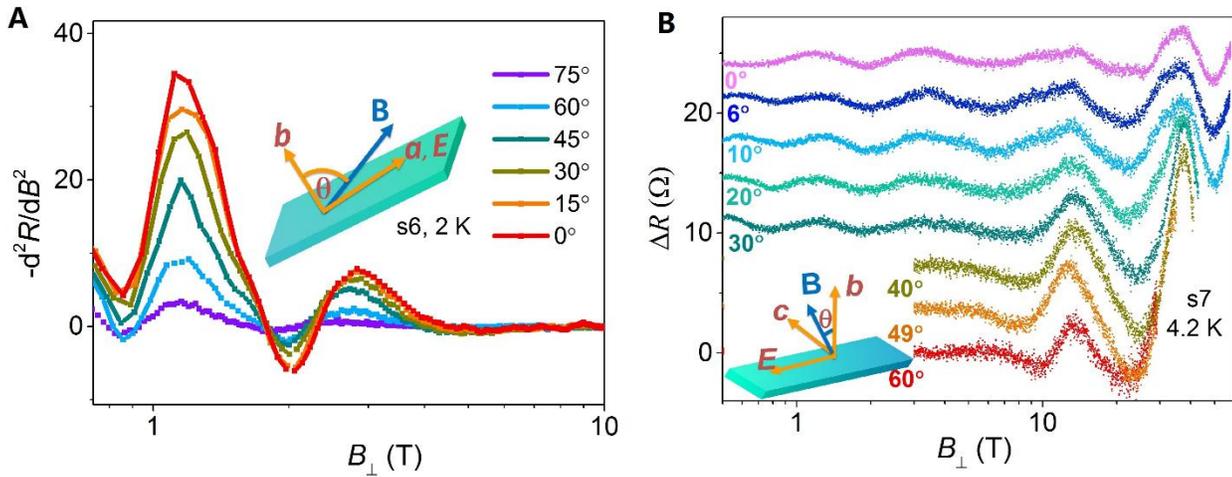

**fig. S7. Strong anisotropy of the bulk ZrTe$_5$.** (**A**) Second derivative of the MR data of s6 vs. the perpendicular component of magnetic field as the sample is rotated in *ab* plane. (**B**) MR oscillations in s7 at 4.2 K vs. the perpendicular component of magnetic field as the sample is rotated in *bc* plane. The 0 °curve indicates a perpendicular magnetic field (*B*//*b* axis).

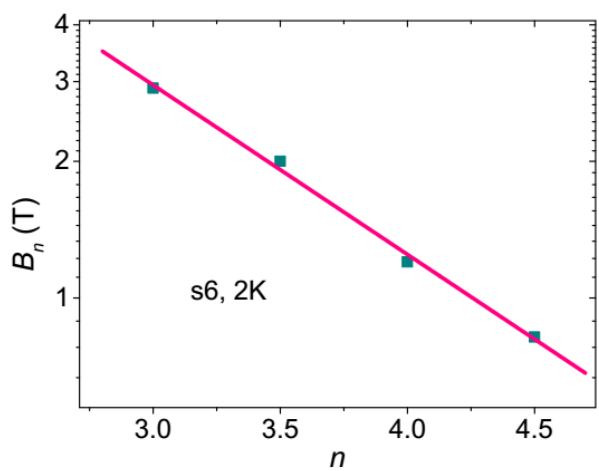

**fig. S8.** Log$B_n$ vs. *n* for the oscillations in s6 measured at the static magnetic field.

## II. Supplemental Discussions on Other Physical Mechanisms

### 1. Exclusion of the Shubnikov-de Haas (SdH) effect

Since our sample shows high mobility, it is expected to observe obvious SdH effect, which occurs at low temperatures in the presence of intense magnetic fields. The SdH effect generally shows a period in $1/B$. By plotting $B_n$ vs. $n$ in a linear scale, however, we find that the oscillations are not periodic in $1/B$ or $B$ (Fig. 4(a) and fig. S5). The Hall traces shown in Fig. 1C indicate that multiple carriers with different carrier densities exist in the $ZrTe_5$ system and thus the superposition of oscillations with different SdH periods may induce exotic oscillations. However, we could not obtain reasonable SdH periods either by performing a FFT analysis for the observed MR oscillations.

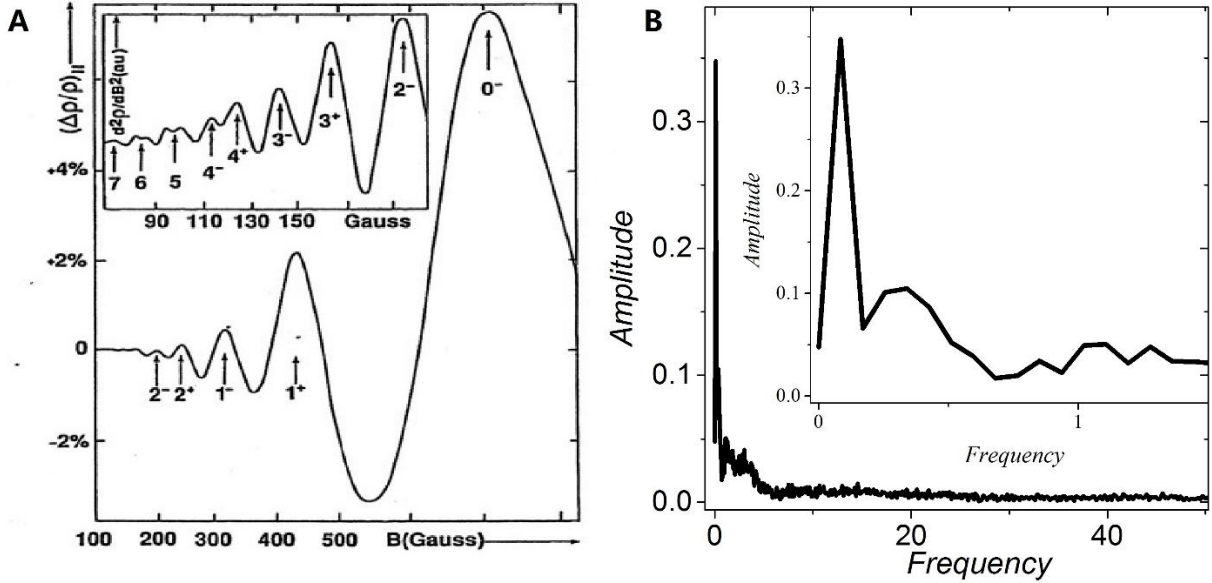

fig. S9. The log-periodic oscillations can not be explained by the SdH effect with Zeeman splitting. (**A**) Typical SdH oscillations with Zeeman splitting (*45*). (**B**) FFT result for the observed oscillations at 4.2 K in the form of $\Delta R$ vs. $1/B$. Inset is the enlargement of the FFT peaks.

Sometimes, the periodic appearance of oscillations under $1/B$ in SdH effect may be broken by spin splitting. The SdH oscillations with Zeeman splitting mainly show two features as follows. (i) In the large B regime, the $n$-th Landau level splits into two peaks generally labelled by $n+$ and $n-$ (fig. S9A) (*45*). Quantitively, one can plot $1/B$ versus $n+$ or $1/B$ versus $n-$, both cases result in a nearly linear dependence on index $n$ for $1/B$ and reasonable Fast Fourier Transform (FFT) results can still be obtained (*23,46-49*). (ii) The Landau levels are split by $g\mu_B B$ for the Zeeman splitting and the spin-splitting parameter is demonstrated as $S$ in the formula $F/B=n+\gamma\pm S/2$. Here $F$ is the SdH period, $n$ is the Landau index, $\gamma$ denotes the phase shift and $S$ is the spin-splitting parameter with $S=\frac{gm^*}{2m_e}$ where $g$ denotes the Landé $g$-factor. The Landé $g$-factor thus can be determined as $g = \frac{2m_e F}{m^*}\left(\frac{1}{B_{n+}} - \frac{1}{B_{n-}}\right)$. The Landé $g$-factor is a constant for a certain material system. Thus, in a plot of the SdH oscillations vs. $1/B$, the spacing of the Zeeman splitting peaks (dips) for the Landau levels should be equidistant.

However, both the above features are not observed in the log-periodic oscillations. (i) The index plot of $1/B_n$ vs $n$ shows no signal of the linear dependence (Fig. 4A). We also perform FFT analysis of the oscillations in the form of $\Delta R$ vs. $1/B$ at 4.2 K, which is shown in fig. S9B. The FFT analysis only gives peak $F$ below 0.3 T, which cannot explain our observed oscillations in the magnetic field beyond 0.3 T. Besides, the peaks below 0.3 T can not be meaningful SdH periods because only one oscillation peak appears below 0.3 T, from which the period cannot be extracted correctly. (ii) The log-periodic oscillations in our observation show no signal of equal spacing for any two groups of peaks or dips (fig. S5), which is different from the trait of the SdH oscillations with Zeeman splitting. In fact, our observed oscillations show the property periodic in log$B$. Thus, the mechanism of the SdH oscillations with Zeeman splitting can be excluded for our observation of the exotic log-periodic oscillations.

## 2. Other Mechanisms beyond the QL

Many physical mechanisms may exist in a three-dimensional (3D) electronic system beyond the QL. In the case of a two-dimensional electron gas, the most famous phenomenon is the fractional quantum Hall effect (FQHE) (*50*), which reflects the fractionalization of an electron. For a long time, similar phenomenon has also been investigated in 3D bulk materials when they are subjected to a sufficiently strong magnetic field. Though the mechanism is still under debate, some MR oscillations in 3D systems beyond the QL had been reported and viewed as precursors to FQHE. For example, the signature of FQHE has been claimed in bismuth (*39*), Bi$_2$Se$_3$ (*40*), graphite (*51*), and ZrTe$_5$ thin flake (*30*). Indeed, the electron-electron interaction becomes crucially important when all the electrons are confined to the lowest Landau level and the FQHE states might be observed in high quality samples (*51*). Trying to understand our observations in this diagram, we assume the first peak observed at ~0.3 T (approaching the estimated ~0.2 T) is at the lowest Landau level. The ratio of the last peak at ~34T (dip at ~50 T) to the first peak at ~0.3 T (dip at ~0.5 T) is as small as 0.01, which corresponds to a fractional index of 1/100, which is rather shocking. Meanwhile, the commonly observed smaller fractional indexes are invisible in the oscillations. Moreover, these oscillations could survive at temperatures as high as 100 K, while the FQHE effect is quite sensitive to a high temperature. Besides, we measured the resistance of s2 vs. sweeping bias at 25 mK under a selectively fixed magnetic field and the results do not agree with the behavior of a typical FQHE state or a Winger crystal state. Thus, it is impossible to assign our observation of the MR oscillations beyond the QL as developing FQHE states.

Theoretically, the electronic spectrum of a three-dimensional electron gas system beyond the QL becomes analogue to a quasi-one-dimensional system. A variety of instabilities may occur in this regime, such as a spin, charge or valley density-wave transition (*38*). The proposed instabilities have been reported in a few materials, such as bismuth (*41*), graphite (*52*), and TaAs (*53*). Very recently, a high magnetic field study work was reported in ZrTe$_5$, in which two anomalous MR peaks in the ultra-quantum regime are explained as the dynamical mass generation at the $n$=1 and $n$=0 Landau levels associated with the density wave transitions (*23*). It is natural, therefore, to consider whether our observations also result from a similar effect. The density-wave transition usually leads a huge increase of the resistance, as observed in other systems (*23,52*). In our observation, the sharp resistance increase is observed from 0 T to about ±1 T. The feature is extending in a wider magnetic field regime than that induced only by the WAL effect, which probably indicates a role of phase transition. However, other oscillations above 3 T do not accompany sharp change of the MR. Besides, the critical magnetic field for the phase transition is generally temperature-dependent (*52*), while the locations in B of oscillations in our samples are almost unchanged at different temperatures. Moreover, the density-wave transition generally occurs at only one critical magnetic field strength, while as many as five oscillating cycles (5 peaks and 5 dips) are observed here. Thus, the mechanism of density-wave transition is not responsible for the oscillations in our observation.

## III. Supplemental Notes on the Theoretical Details

Considering that ZrTe$_5$ can be a Dirac semimetal (*25,29*), we treat the light hole as a massless Dirac fermion, and solve the spectrum of the massless particle in a Coulomb attraction potential from the charge center. In the following part, we consider the charge number $Z = 1$. Because of the small carrier density of holes ($2.6 \times 10^{15}$ cm$^{-3}$), the screening effect can be neglected. Thus, we consider the solution of the Dirac equation for a hole in a Coulomb potential and under the magnetic field as

$$\begin{bmatrix} m & \hbar v_F \left( \vec{\sigma} \cdot \vec{k} + \frac{e}{\hbar c} \vec{\sigma} \cdot \vec{A} \right) \\ \hbar v_F \left( \vec{\sigma} \cdot \vec{k} + \frac{e}{\hbar c} \vec{\sigma} \cdot \vec{A} \right) & -m \end{bmatrix} \begin{bmatrix} \Psi_1 \\ \Psi_2 \end{bmatrix} = [E - V(\vec{R})] \begin{bmatrix} \Psi_1 \\ \Psi_2 \end{bmatrix} \quad \text{(SA1)}.$$

We consider the symmetry gauge $\vec{A} = \frac{B}{2}(-y, x, 0)$, and the Coulomb attraction of central charge reads $V(\vec{R}) = \frac{-e^2}{4\pi\varepsilon_0 R}$. The solution can be represented by:

$$\begin{bmatrix} \Psi_1(\vec{R}) \\ \Psi_2(\vec{R}) \end{bmatrix} = \begin{bmatrix} Y^{jm}_{j-1/2}(\theta, \varphi) u_1(R)/R \\ i Y^{jm}_{j+1/2}(\theta, \varphi) u_2(R)/R \end{bmatrix} \quad \text{(SA2)}.$$

Here, we use the spinor function $i Y^{jm}_{j\pm 1/2}(\theta, \varphi)$ with J. J. Sakurai's notation (*54*), and we define $\kappa = \pm(j + 1/2)$ with $j$ being a positive half integer for the following usage.

**1. Discrete scale invariance (DSI) without magnetic field**

After eliminating the angular wave function, we have the radial equation with the fine structure constant defined by $\alpha = \frac{e^2}{4\pi\varepsilon_0 \hbar v_F}$:

$$\frac{d}{dR} \begin{bmatrix} u_1(R) \\ u_2(R) \end{bmatrix} = \begin{bmatrix} \frac{\kappa}{R} & \frac{E+m}{\hbar v_F} + \frac{\alpha}{R} \\ -\left(\frac{E-m}{\hbar v_F} + \frac{\alpha}{R}\right) & -\frac{\kappa}{R} \end{bmatrix} \begin{bmatrix} u_1(R) \\ u_2(R) \end{bmatrix} \quad \text{(SA3)}.$$

The system has scale invariance for the massless Dirac equation ($m = 0$). Based on WKB analysis, we obtain the simplified differential equation for $u_i$ with $i = 1, 2$:

$$\frac{d^2}{dR^2} u_i(R) = \frac{\kappa^2}{R^2} u_i(R) - \left(\frac{E}{\hbar v_F} + \frac{\alpha}{R}\right)^2 u_i(R) \quad \text{(SA4)}.$$

The exact differential equation deduced from equation (SA3) without WKB approximation also have term of order $O(R^{-1})$ (*12-14*) and can be neglected in the small $R$ limit. Thus, the radial momentum satisfies:

$$p_R^2 = \hbar^2 \left[ \left(\frac{E}{\hbar v_F} + \frac{\alpha}{R}\right)^2 - \frac{\kappa^2}{R^2} \right] \quad \text{(SA5)}.$$

For $\alpha < |\kappa|$, namely the subcritical regime, the system cannot have bound states due to Klein tunneling (*12*). In contrast, for $\alpha > |\kappa|$ namely the supercritical regime, the system can possess quasi-bound states (*23*). This phenomenon can also exist in two-dimensional graphene system (*13,14*). The radial momentum can only have real solution under the supercritical collapse case[23]:

$$s_0^2 = (\alpha)^2 - \kappa^2 > 0 \quad \text{(SA6)}.$$

In the following, we focus on the lowest angular momentum channel with $\kappa = \pm 1$. The radial momentum can be written as $p_r \approx \frac{\hbar s_0}{r}$. Based on the WKB (Wentzel–Kramers–Brillouin) analysis (*55*), we can obtain the energy spectrum of the two-body quasi-bound states. To be specific, the semi-classical quantization $\int_{R_0}^{R_n} p \cdot dr = n\pi\hbar$ results in DSI for the quasi-bound states with the radius satisfying:

$$R_n/R_{n-1} = e^{\pi/s_0} \quad \text{(SA7)}.$$

The quasi-bound state energy satisfies the relation with a lattice cutoff $R_0$:

$$E_n = \frac{\hbar v_F \alpha}{R_n} = \frac{\hbar v_F \alpha}{R_0} e^{-n\pi/s_0} \quad \text{(SA8)}.$$

Here, we neglect the spectra width of these quasi-bound states originated from the Klein tunneling, which cannot influence the energy ratio of these quasi-bound states. Specifically, for the case of Dirac bands in ZrTe$_5$, the Fermi velocity $v_F \approx 4.0 \times 10^5$ m/s (*25,29,31*), and thus the supercritical collapse condition $\alpha > 1$ can be satisfied (with $\alpha \approx 5.50$). Considering the experimental value of $\alpha \approx 5.50$, there exist other angular momentum channels satisfying the supercritical condition (SA6) with different value of $s_0$. For angular momentum channel $\kappa = \pm 1, \pm 2, \pm 3$, the corresponding values of $e^{2\pi/s_0}$ locate in the relatively narrow range of [3.20, 4.21], and $e^{\pi/s_0}$ for $\kappa = \pm 4, \pm 5$ deviates from this regime.

## 2. Approximate DSI and dissolution of quasi-bound states under magnetic field

The magnetic field introduces a new characteristic length $l_B = \sqrt{\hbar c/eB}$. We consider the case that the Coulomb attraction $|V(R_n)| = \frac{\alpha \hbar v_F}{R_n}$ is much larger than the Landau level spacing $E_B = \frac{\sqrt{2}\hbar v_F}{l_B}$, which is identical to $R_n \ll \frac{\sqrt{2}}{2}\alpha l_B$. Expanding in spinor spherical harmonic function, the radial equation of massless Dirac particle reduces to:

$$\frac{d}{dR}\begin{bmatrix} u_1(R) \\ u_2(R) \end{bmatrix} = \begin{bmatrix} \frac{\kappa}{R} + \frac{R}{2l_B^2} & \frac{E}{\hbar v_F} + \frac{\alpha}{R} \\ -\left(\frac{E}{\hbar v_F} + \frac{\alpha}{R}\right) & -\left(\frac{\kappa}{R} + \frac{R}{2l_B^2}\right) \end{bmatrix} \begin{bmatrix} u_1(R) \\ u_2(R) \end{bmatrix} \quad (SA9).$$

The radial equation can be approximately written as:

$$\frac{d^2}{dR^2} u_i(R) = \left(\frac{\kappa}{R} + \frac{R}{2l_B^2}\right)^2 u_i(R) - \left(\frac{E}{\hbar v_F} + \frac{Z\alpha}{R}\right)^2 u_i(R) \quad (SA10).$$

We can find that the influence of the magnetic field truly breaks the scale invariance of the system. The radial momentum satisfies $p_R^2 \approx \left(\frac{Z\alpha}{R}\right)^2 - \left(\frac{\kappa}{R} + \frac{R}{2l_B^2}\right)^2$, and the radial momentum has real solution for $R < R_c$ with $R_c^2 = l_B^2 \frac{2s_0^2}{\sqrt{s_0^2 + \kappa^2} + \kappa} \approx 2s_0 l_B^2$. In the regime $R < R_c$, the radial momentum can be written as:

$$p_R \approx \frac{\hbar}{R}\sqrt{(Z\alpha)^2 - \left(\kappa + \frac{R^2}{2l_B^2}\right)^2} \approx \frac{\hbar s_0}{R}\left(1 - \frac{R^2}{2s_0^2 l_B^2}\right) \quad (SA11).$$

The semi-classical quantization condition $\int_{R_0}^{R_n} p \cdot dr = n\pi\hbar$ results in the approximate DSI:

$$\ln\frac{R_n}{R_{n-1}} = \frac{\pi}{s_0} + \frac{R_n^2}{4s_0^2 l_{Bn}^2} - \frac{R_{n-1}^2}{4s_0^2 l_{Bn-1}^2} \in \left(\frac{\pi}{s_0} - \frac{1}{2s_0}, \frac{\pi}{s_0} + \frac{1}{2s_0}\right) \quad (SA12).$$

Thus, the corresponding magnetic field strength satisfies the approximate DSI:

$$\ln\frac{B_{n+1}}{B_n} \in \left(-\frac{2\pi}{s_0} - \frac{1}{s_0}, -\frac{2\pi}{s_0} + \frac{1}{s_0}\right) \quad (SA13).$$

## 3. Comparison to the experimental results in ZrTe$_5$

We consider the Fermi velocity $v_F \approx 4.0 \times 10^5$ m/s for the Dirac bands in ZrTe$_5$ (*25,29,31*), and then $Z\alpha \approx 5.50$ gives the scaling factor $s_0 = \sqrt{(Z\alpha)^2 - 1} \approx 5.4$. The influence of the magnetic field on the quasi-bound states in the Dirac system can be two-fold.

The magnetic field induces a harmonic trap in the plane perpendicular to the magnetic field, and breaks the two-body quasi-bounds with large radius $R_n > \sqrt{2s_0}l_B$. This breakdown of the two-body bounds states induces change to the mobile carrier density, and thus gives rise to the log-periodic MR oscillations. For angular momentum channel with $\kappa = 1$, the theoretically estimated value for the approximate DSI factor $\lambda$ locates in the range of (2.76, 4.06). The approximate DSI feature is further confirmed by numerical simulation of equation (Fig. 4C), and the approximate DSI with the estimated $\lambda \in (3.35, 3.50)$. Considering the contribution from higher angular momentum channels, the values of approximate DSI locate in a broader range, which is consistent with the experimental results shown in the main text. Based on the WKB analysis (*55*), we have obtained the energy spectrum evolution of the quasi-bound states under the magnetic field. When enlarging the magnetic field, the energy of the *n*-th quasi-bound states approaches the Fermi energy at the magnetic field $B_n$. Elastic scattering between the mobile carriers and the quasi-bound states around the Fermi energy strongly influences the transport property. Moreover, the scattering between mobile carriers and Anderson impurity also contributes to the resistance, which manifests as the background of the MR oscillations.

### A.3.1 The total magneto-conductivity

Based on the t-matrix approximation, we derive the longitudinal conductivity beyond the quantum limit under large magnetic fields (the details are given in the theoretical preprint Haiwen Liu *et. al.*):

$$\sigma_{xx}(\varepsilon_F) = \frac{4e^2}{h} l_B^2 \left( n_s + n_c \frac{t^2}{8\pi \cdot \hbar v_F l_*^{-1} \cdot \Gamma(B)} \sum_n \frac{\Gamma(B)^2}{(\varepsilon_F - \varepsilon_n(B))^2 + \Gamma(B)^2} \right) \quad \text{(SA14)}.$$

Here $n_s$ is the density of short-range scatterers, $n_c$ is the density of Coulomb scatterers, $l_*$ is the effective length along the magnetic field and $t$ is the coupling strength between the bound states with the continuum of the lowest Landau level, $\varepsilon_F$ is the Fermi energy, $\varepsilon_n(B)$ is the energy for the $n$-th bound state and $\Gamma(B) \propto \sqrt{B}$ is the width mainly determined by the broadening effect of the lowest Landau level. The above microscopic formula can be further simplified into an empirical form which are more suitable for fitting the experimental data:

$$\sigma_{xx}(\varepsilon_F) = = \frac{4e^2}{h} l_B^2 \left( n_s + n_c \frac{t^2}{8\pi \cdot \hbar v_F l_*^{-1} \cdot \Gamma(B)} \frac{\eta^2}{\sin^2\left(\frac{s_0}{2} \ln\left(\frac{B}{B_0}\right)\right) + \eta^2} \right) \quad \text{(SA15)},$$

$$\sigma_{xy} = \frac{2\pi e^2}{h} l_B^2 \cdot N \quad \text{(SA16)},$$

Here, N denotes the total carrier density, $\eta$, $s_0$ and $B_0$ are fitting parameters. In equation (SA15), the first term denotes the Anderson impurity scattering of the mobile carriers, which leads to linear-B dependent MR previously obtained by A. A. Abrikosov (*56*); and the second term denotes the resonant scattering between the mobile carriers and the quasi-bound states, which gives rise to log-periodic correction to the MR.

### A.3.2 The DSI feature in MR and the fitting parameters of log-periodic oscillations.

From the magnetoconductivity in equation (SA15) and (SA16), we can obtain longitudinal MR by relation $\rho_{xx} \approx \frac{\sigma_{xx}}{\sigma_{xy}^2}$. Here, we have considered the property of our ZrTe$_5$ sample that $\rho_{xx} \ll \rho_{xy}$ and subsequently $\sigma_{xx} \ll \sigma_{xy}$ for B > 0.5 Tesla, and we use the relation $\rho_{xx} = \frac{\sigma_{xx}}{\sigma_{xx}^2 + \sigma_{xy}^2}, \rho_{xy} = \frac{-\sigma_{xy}}{\sigma_{xx}^2 + \sigma_{xy}^2}$. Thus, the longitudinal MR $\rho_{xx}$ reads:

$$\rho_{xx} = \frac{h}{\pi^2 \cdot e^2} l_B^{-2} \left( \frac{n_s}{N^2} + \frac{n_c}{N^2} \frac{t^2}{8\pi \cdot \hbar v_F l_*^{-1} \cdot \Gamma(B)} \frac{\eta^2}{\sin^2\left(\frac{s_0}{2} \ln\left(\frac{B}{B_0}\right)\right) + \eta^2} \right) \quad \text{(SA17)}.$$

In solid state systems, there exist multiple mechanisms of scattering, which is different from the ultra-cold atomic systems. The resonant scattering of the two-body quasi-bound states contribute to the log-periodic oscillations as shown in the second term in equation (SA16), while the scattering of Anderson impurity gives the linear-B dependent MR as a background. Although the background due to scattering from Anderson impurity truly breaks the DSI feature, the DSI feature recovers after subtracting the non-oscillating background. This procedure is widely utilized in the analysis of quantum oscillations in solid state system, such as *Shubnikov–de Haas* oscillations and *de Haas-van Alphen* oscillations (*16*). As shown in the Fig. 2C and Fig. 3C in the main text, the data obeys the DSI properties after substation of non-oscillating background. Moreover, for *Shubnikov–de Haas* oscillations, one commonly uses the notation periodic in 1/B to highlight the underneath Landau level physics, although the oscillation amplitude changes with B (*16*)). For similar reason, we use the notation of log-periodic oscillations in the title to highlight the effect of quasi-bound states with DSI, although the oscillation amplitude also changes with B.

The second term in equation (SA17) satisfies the DSI with envelop function proportional to $\sqrt{B}$, and leads to the log-periodic corrections. The MR oscillation data shown in Fig. 2C can be well fitted by the theoretical results in equation (SA17), which is shown in Fig. 4D in the main text. Based on the fitting parameters shown in Table S3, an averaging value of $s_0 \approx 5.4$ is obtained and thus the Fermi velocity $v_F \approx 4.0\times10^5$ m/s can be deduced for the Dirac bands in ZrTe$_5$, which is very close to the results in previous literature (*25, 29, 31*). Besides, the log-periodic oscillations at different temperatures can also be reproduced by the theoretical formula (fig. S10) and the fitting parameters are shown in Table S4. The factor $s_0 \approx 5.2$ doesn't vary at different temperatures, while the parameter $\eta^2$ increases with the increasing temperature due to the thermal broadening effect. Thus, based on our quantitative analysis, the log-periodic oscillating MR in ZrTe$_5$ beyond the quantum limit obeys the approximate DSI feature, which originates from the resonant scattering between the mobile carriers and the two-body quasi-bound states around the Fermi energy.

Table S3 Fitting parameters of log-periodic oscillations in different samples.

| Samples / Parameters | s6 | s7 | s9 |
|---|---|---|---|
| $s_0$ | 5.190 | 5.501 | 5.620 |
| Standard error of $s_0$ | 0.005 | 0.009 | 0.008 |
| $\eta^2$ | 0.492 | 0.466 | 0.785 |
| Standard error of $\eta^2$ | 0.008 | 0.014 | 0.015 |
| $B_0$(4.2 K) (T) | 0.267 | 0.361 | 0.394 |
| Standard error of $B_0$ (T) | 0.001 | 0.002 | 0.002 |
| c | 0.583 | 0.558 | 0.636 |
| Standard error of c | 0.003 | 0.006 | 0.004 |

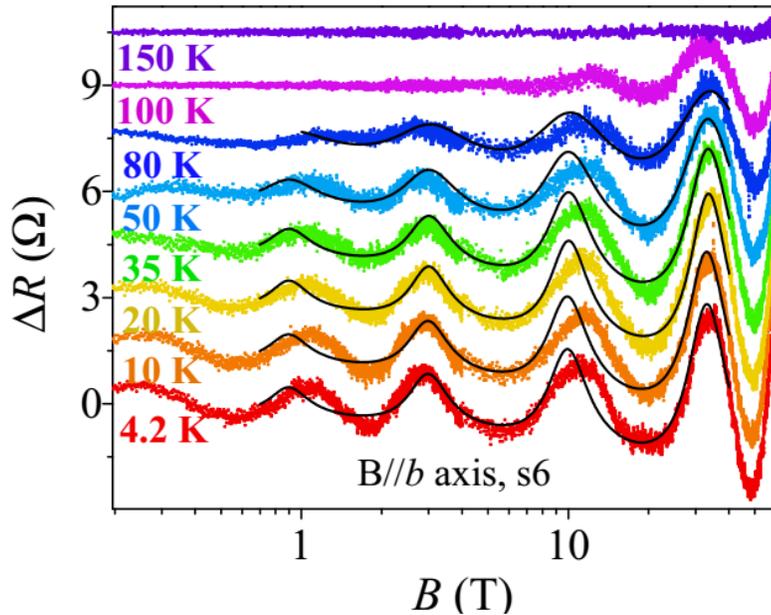

fig. S10. The observed log-periodic oscillations at different temperatures after the subtraction of background. The black lines are the theoretical fitting by fixing $B_0$ at 0.267.

Table S4 Fitting parameters of log-periodic oscillations in s6 at different temperatures.

| Temperature / Parameters | 4.2 K | 10 K | 20 K | 35 K | 50 K | 80 K |
|---|---|---|---|---|---|---|
| $s_0$ | 5.222 | 5.225 | 5.207 | 5.210 | 5.215 | 5.200 |
| Standard error of $s_0$ | 0.002 | 0.002 | 0.001 | 0.001 | 0.001 | 0.002 |
| $\eta^2$ | 0.341 | 0.353 | 0.319 | 0.403 | 0.736 | 1.777 |
| Standard error of $\eta^2$ | 0.011 | 0.011 | 0.008 | 0.009 | 0.011 | 0.021 |
| c | 0.509 | 0.512 | 0.492 | 0.531 | 0.645 | 0.770 |
| Standard error of c | 0.007 | 0.006 | 0.005 | 0.004 | 0.003 | 0.002 |

### A.3.3 Zeeman effect at large magnetic fields

The Zeeman effect needs to be considered if it is comparable to the Fermi energy. Based on the previous analysis in ZrTe$_5$ (31), the Zeeman spin splitting can show two kinds of influence on the lowest Landau level.

Case (a): This case happens when magnetic field is applied along the crystallographic b-axis (31). The b-axis is perpendicular to the layered ZrTe$_5$ sample. The Zeeman field gives rise to spin-polarized conduction band and valence band with energy $E_0 = \pm\sqrt{(hv_F k_\parallel)^2 + \left(\frac{g}{2}\mu_B B\right)^2}$. The influence of Zeeman energy can be represented by the effective Fermi energy:

$$\varepsilon_F = \sqrt{\varepsilon_{F0}^2 + \left(\frac{g}{2}\mu_B B\right)^2} \tag{SA18}$$

Based on the carrier density information from the Hall data, we estimate the Fermi energy in our ZrTe$_5$ samples that is about $\varepsilon_{F0}$ = 15 meV. Considering $g \approx 20$ (31,35), the Zeeman energy is comparable to the Fermi energy at about 30 T.

Case (b): The Zeeman field only shifts the wave vector parallel to the magnetic field and the final energy spectrum of the lowest Landau level reads: $E_0 = \pm hv_F k_\parallel - \frac{g}{2}\mu_B B$. For this case, the Zeeman effect doesn't influence the Fermi energy or our results. This case happens when magnetic field is applied along the c-axis (31).

In the following, we consider the quantitative influence of the Zeeman effect on our results for case (a), since case (a) happens when the magnetic field is applied along b-axis just like our experimental measurement setup.

Based on the relation of $\sigma_{xy}(\varepsilon_F) = \frac{4e^2}{h}l_B^2 \cdot N$ ($N$ denoting the total carrier density), we obtain the magneto-resistivity $\rho_{xx} \approx \sigma_{xx} \cdot \sigma_{xy}^{-2}$ (here we used the property $\sigma_{xx} \ll \sigma_{xy}$ for our data). By setting $\varepsilon_F = 0$, equation (SA14) is well consistent with equation (SA15), and the fitting curves are shown in fig. S11A. As a real material system, the ZrTe$_5$ has a small Fermi energy $\varepsilon_{F0}$ = 15 meV, and the Zeeman effect needs to be considered as shown in equation (SA18). After considering the Zeeman effect, the microscopic result from equation (SA14) with Zeeman effect shows a better match for the experimental results (s6), although there is no significant influence on the fitting curves (fig. S11B). Here we only consider the Coulomb scatterers, and neglect the short-range scatterers which only contribute a non-oscillating background.

We have to admit that the fitting shown in fig. S11B still deviates from the experimental data in both the small field limit and the large field limit. This deviation can be attributed to the background subtraction near the boundary. Because the subtracted background at the both ending fields is largely affected by the boundary condition, the extracted characteristic magnetic field $B_n$ for the oscillations at the boundary magnetic fields show a small error, which would lead to the certain deviations between the theoretical fit and the experimental data.

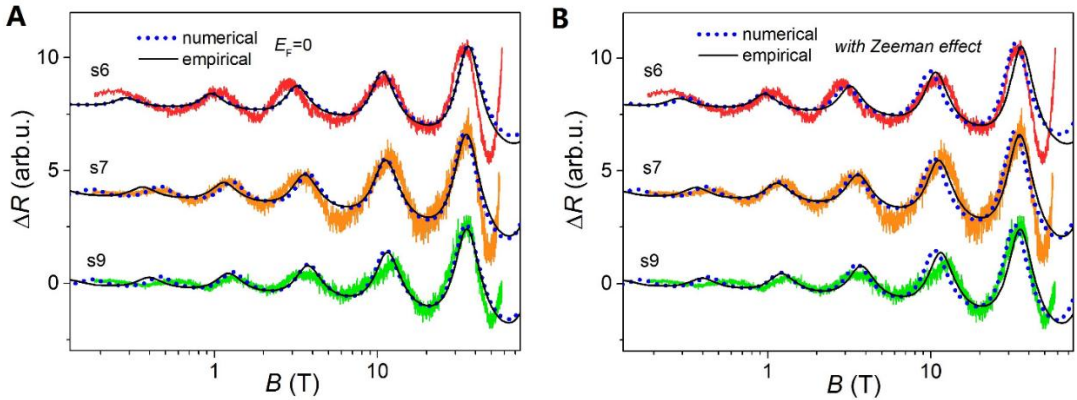

fig. S11 The fitting of experimental data based on the microscopic numerical calculation and the empirical formula (**A**) without and (**B**) with the consideration of the Zeeman effect.

## 4. Related Discussions
### 4.1 Mass influence on the DSI

Based on the aforementioned analysis, we obtain the DSI of the two-body quasi-bound states in the supercritical collapse of Dirac particles with Coulomb attraction. The magnetic field introduces a new length scale into the system, and breaks the DSI down to approximate DSI. Here, we consider the influence of mass term, which introduces a new energy scale and can also break the DSI down to approximate DSI for certain quasi-bound states.

We consider the magnetic field case with $R_n \ll \frac{\sqrt{2}}{2} \alpha l_B$. The system is still approximate spherical, and after expanding in spinor spherical harmonic function one can obtain the radial equation with mass term $m$:

$$\frac{d^2}{dR^2} u_i(R) = \left(\frac{\kappa}{R} + \frac{R}{2l_B^2}\right)^2 u_i(R) - \left[\left(\frac{E}{\hbar v_F} + \frac{\alpha}{R}\right)^2 - \frac{m^2}{(\hbar v_F)^2}\right] u_i(R) \quad (SA19).$$

For small mass $m < \frac{\alpha \hbar v_F}{R_n}$ and $R_n < \sqrt{2s_0} l_B$, we can still obtain bound states with the momentum satisfying $p_R \approx \frac{\hbar}{R}\sqrt{\alpha^2 - \left(\kappa + \frac{R^2}{2l_B^2}\right)^2 - \frac{m^2 R^2}{(\hbar v_F)^2}}$. The semi-classical quantization condition $\int_{R_0}^{R_n} p_R \cdot dR = n\pi\hbar$ gives the approximate DSI for the two-body quasi-bound states:

$$\ln\frac{R_n}{R_{n-1}} \in \left(\frac{\pi}{s_0} - \frac{1}{2s_0}(1 + \chi^2), \frac{\pi}{s_0} + \frac{1}{2s_0}(1 + \chi^2)\right) \quad (SA20),$$

with $\chi = \frac{m l_B}{\hbar v_F}$. Thus, the mass term has little influence on the DSI under large magnetic field limit with $m \ll \frac{\hbar v_F}{l_B}$. The typical value of $\frac{\hbar v_F}{l_B}$ is about 12 meV at 1 Tesla.

Other physical mechanisms may also influence the DSI feature, which need further theoretical analysis. Firstly, the screening effect from the continuous band gives rise to a long-range cut-off for the two-body bound states, and the lattice constant gives the short-range cutoff. Based on the carrier density, estimation of the range with DSI feature locates in 0.4-60 nm, and the corresponding magnetic field locates in range of 0.2-150 T. Moreover, the continuous bands also contribute to the magnetoresistance, which is rendered as the envelope function of magnetoresistance without the log-periodic oscillations.

### 4.2 Relation to Efimov states

After completion of this work, we became aware of a preprint, which addresses the quasi-bound states and the signature of broken continuous scale symmetry in graphene probed by scanning tunneling microscope (*43*). Here we give a comparison of our results to those of graphene system with charge vacancy. The underlying physics of the DSI in our system is similar to the graphene case (*13*). The difference is that our work focuses on the three-dimensional Dirac fermions with Coulomb attraction. More importantly, the clear evolution of the two-body quasi-bound state under magnetic field is elucidated, and the remarkable influence of those quasi-bound states on the magnetotransport properties is revealed. In the experimental aspect, we have observed as many as five oscillating cycles (5 peaks and 5 dips) to clearly demonstrate the DSI and the two-body quasi-bound states in the condensed matter systems. Our work also indicates that the intriguing log-periodic oscillations (DSI) are potentially universal in the topological materials with Coulomb attraction.

Moreover, although the DSI in our experiments remind us of the property of Efimov trimer states. However, we have found that the prerequisite for the Efimov trimer states is hard to be met in our system. The reason is twofold. Firstly, the resonant scattering condition for Efimov trimer states is hardly matched under the change of magnetic field in large regime. Secondly, the Efimov trimers are composite states of Schrödinger particles, while in our system the massless Weyl particles play a major role. Thus, the two-body quasi-bound states can be responsible for our observation of the novel phase with the DSI by showing the log-periodic quantum oscillations on the MR.